\documentclass[prl,superscriptaddress,amsfonts,amssymb,amsmath,floats,twocolumn,aps,footinbib,shownopacs]{revtex4-2}
\usepackage[pdftex]{graphicx}
\usepackage{subfigure}
\usepackage{dsfont}
\usepackage{lipsum}
\usepackage{dcolumn}
\usepackage{bm}
\usepackage{color}
\usepackage{physics}
\usepackage{multirow}
\usepackage[pdftex,colorlinks=true, linkcolor=blue,citecolor=blue,filecolor=blue]{hyperref}
\usepackage{cancel}
\usepackage{xcolor}

\begin{document}
\title{Quantum Thermalization via Travelling Waves}
\newcommand{\mygraphicwidth}{1.0\textwidth}

\author{Antonio Picano} 
\affiliation{JEIP, UAR 3573, CNRS, Coll\`ege de France, PSL Research University, 11 Place Marcelin Berthelot, 75321 Paris Cedex 5, France}
\author{Giulio Biroli}
\affiliation{Laboratoire de Physique de l'\'Ecole Normale Sup\'erieure, ENS, Universit\'e PSL, CNRS, Sorbonne Universit\'e, Universit\'e Paris Cit\'e,
F-75005 Paris, France}
\author{Marco Schir\`o} 
\affiliation{JEIP, UAR 3573, CNRS, Coll\`ege de France, PSL Research University, 11 Place Marcelin Berthelot, 75321 Paris Cedex 5, France}

\begin{abstract}
Isolated quantum many-body systems which thermalize under their own dynamics are expected to act as their own thermal baths, thereby loosing memory of initial conditions and bringing their local subsystems to thermal equilibrium. Here we show that the infinite-dimensional limit of a quantum lattice model, as described by Dynamical Mean-Field theory (DMFT), provides a natural framework to understand this self-consistent thermalization process. 
Using the Fermi-Hubbard model as working example, we demonstrate that the emergence of a self-consistent bath occurs via a sharp thermalization front, moving balistically and separating the initial condition from the long-time thermal fixed point.  We characterize the full DMFT dynamics through an effective temperature for which we derive a travelling-wave equation of the Fisher-Kolmogorov-Petrovsky-Piskunov (FKPP) type. This equation allows to predict the asymptotic shape of the front and its velocity, which match perfectly the full DMFT numerics. Our results provide a new angle to understand the onset of quantum thermalisation in closed isolated systems.
\end{abstract}
\maketitle 	

\emph{Introduction - } Understanding how isolated quantum many-body systems approach thermal equilibrium is at the same time of fundamental interest for the foundation of quantum statistical mechanics~\cite{polkovnikov2011colloquium,Gogolin_2016,dalessio2016from}, as well as experimentally relevant for many-platforms for quantum simulation~\cite{kaufman2016quantum,kranzl2023experimental}.
Much of our insight on quantum thermalization relies on the  Eigenstates Thermalization Hypothesis (ETH)~\cite{MVBerry_1977,Deutsch91,Srednicki_ETH,MarkSrednicki_1999}, 
according to which thermalization is encoded in the eigenstates of non-integrable quantum many-body systems, i.e., the expectation value of local few-body observables computed on such eigenstates coincides with the prediction of statistical mechanics in the appropriate ensemble~\cite{rigol2008thermalization,Biroli_Corinna_prl10,ikeda2013finite,kim2014testing,reiman2016,hallam2019thelyapunov,foini2019eigenstate,pappalardi2022eigenstate,chan2022manybody}. This ansatz allows to reconcile with the paradox of quantum thermalization and the loss 
of memory of initial condition under unitary dynamics. 

A more qualitative picture of quantum thermalization, often evoked but to the best of our knowledge never fully discussed in detail, is based on the idea that a thermalizing system should be able to \emph{act as its own thermal bath}~\cite{Nandkishore2015}, allowing it to thermalize small subsystems within itself and to loose memory of initial condition. Specifically, this means that if one focuses on finite-sized subsystems within a thermodynamically large many-body system and 
computes the dynamics of their reduced density matrix after tracing the rest of the system out, this would equilibrate to the appropriate statistical mechanics ensemble prediction. This idea, although physically more transparent and intuitively motivated, raises also a number of intriguing questions. 

In particular, one can wonder how this self-consistent thermalization should work in practice and what should be the key features of this emergent self-consistent environment. Indeed, the conventional statistical mechanics~\cite{LandauStatPhys} or open-quantum system~\cite{breuer2002theory} picture for a system coupled to a bath usually assumes a rather different regime, in which the bath is fixed and infinite and the small system equilibrates with it, thus losing memory of its initial state.  On the other hand, thermalization in isolated systems  implies the conversion of excitation energy into heating, a process
which requires the bath to update itself. Recently, motivated by this perspective, there have been attempts to revisit the concepts of thermal baths and standard textbook results such as Fermi Golden Rule~\cite{micklitz2022emergence}.

\begin{figure*}
 	\centering \includegraphics[width=\mygraphicwidth]
 {Figure01U2.png}
	\caption{\label{fig:DMFT_confronto} {\bf Thermalization in Infinite Dimensions with DMFT}: Time evolution of the distribution function (\textbf{a}) and spectral function (\textbf{b}) after an energy excitation, for $U=2$ (initial inverse temperature $\beta_i=25$), leading to a thermal equilibrium fixed point. Full self-consistent DMFT solution ($n=\infty$ in the notation of panels (\textbf{c})-(\textbf{d})).  {\bf Thermalization Front}: time evolution of the total energy (\textbf{c}) and double occupation (\textbf{d}) after an energy excitation, for different DMFT iteration number (coloured lines from blue to red as a function of the DMFT iteration number $n \in [0,99]$) and compared to the full self-consistent DMFT solution (black line). Emergence of a thermalization front in both energy and double occupation, panels (\textbf{e})-(\textbf{f})}.
\end{figure*}
In this Letter, we show how discussing the problem of quantum thermalization in the limit of infinite dimensions naturally provides a theoretical framework to understand the emergence of a self-consistent quantum bath. We focus on strongly interacting lattice fermions which in this limit are mapped exactly through 
Dynamical Mean-Field Theory (DMFT) onto a small interacting quantum system, an impurity, embedded in a self-consistent non-equilibrium environment~\cite{Georges1996,Aoki2014}.

By solving the quantum dynamics of this impurity model and following the self-consistency as it unravels, we show that thermalization emerges in DMFT through a sharp \emph{thermalisation front} in local observables, distinguishing between past (initial condition) and future (thermal equilibrium) and moving at constant velocity. We characterize the DMFT dynamics in terms of an effective temperature for which we derive a traveling wave equation of the Fisher-Kolmogorov-Petrovsky-Piskunov (FKPP) type~\cite{brunetderrida2015}, recently used to describe scrambling and chaos~\cite{ALEINER2016378,xu2019locality,aron2023traveling,aron2023kinetics} and that we here demonstrate to capture the onset of quantum thermalisation. Our analytical results for the front shape and velocity match those of the full DMFT numerical simulation.

%
\emph{Thermalization in Infinite Dimension and DMFT - }
We begin by reviewing the large connectivity limit of fermionic many-body systems given by non-equilibrium extensions of DMFT~\cite{Georges1996,Aoki2014}. We focus on the Fermi-Hubbard model whose Hamiltonian reads
\begin{align}
	\label{ham:hubb}
	\hat H=  -\frac{t_h}{\sqrt{z}} \sum_{\langle i,j\rangle,\sigma}c^{\dagger}_{i \sigma}c_{j\sigma} + U\sum_{j} \big(\hat{n}_{j\uparrow}-\tfrac12\big)\big(\hat{n}_{j\downarrow}-\tfrac12\big)\,.
\end{align}
Here, $c_{j,\sigma}$ annihilates a fermion with spin $\sigma \in \{\uparrow,\downarrow\}$  at lattice site $j$, $\hat{n}_{j\sigma}=c^{\dagger}_{j\sigma}c_{j\sigma}$ is the particle density, $t_h$ the hopping between site $i$ and its $z$ neighbors, and $U$ the on-site interaction strength.  In the $z\rightarrow\infty$ limit, 
the Hubbard model maps onto
a self-consistent quantum impurity model, describing a single site coupled to a nonequilibrium bath, whose effective action on the Keldysh contour $\mathcal{C}$ reads
\begin{align}\label{eqn:qim}
	\mathcal{S}
	=\mathcal{S}_{\text{loc}}-i
	\int_\mathcal{C} dt dt' 
	\sum_\sigma
	c_\sigma^\dagger(t) \Delta_\sigma(t,t')c_\sigma(t')\,.
\end{align}
Here $\mathcal{S}_{\text{loc}}$ is the local (on-site) part of the action while the second term describes the coupling to the self-consistent bath mimicking the effect of the rest of the lattice. The spectrum and occupation of the bath are encoded in the hybridization function $\Delta_{\sigma}(t,t')$ which is given, through the DMFT self-consistency relation, by a function of the local Green's functions of the impurity, $G_{\sigma}(t,t')=-i\langle T_{\mathcal{C}} c_{\sigma}(t)c^{\dagger}_{\sigma}(t')\rangle$.  In the following, we consider a Bethe lattice, with hopping $t_h=1$ setting the unit of energy and semi-elliptic density of states $D(\epsilon)=\sqrt{4-\epsilon^2}/(2\pi)$, for which the self-consistency relation simply reads $\Delta_{\sigma}(t,t')=t_h^2 G_{\sigma}(t,t')$. Solving the dynamics of the Hubbard model within DMFT amounts therefore to solve the dynamics of the quantum impurity model in Eq.~(\ref{eqn:qim}) and implement the self-consistency condition until convergence. We employ the self-consistent Second-order Perturbation Theory (SPT) as an impurity solver~\cite{supplementary}. Being a conserving approximation, it is particularly suited to studying thermalization in isolated quantum systems and the energy transfer between the impurity and the bath at weak coupling~\cite{Aoki2014}. 
For the rest of the paper we fix $U=2$, even though our results are qualitatively similar for other values of the interaction in the weak-coupling regime~\cite{supplementary}.
We fix the initial inverse temperature to $\beta_i=25$ and analyze the dynamics induced by a photo-excitation. This is in practice implemented by coupling the system, for a short interval of time, to a suitable bath which removes low-energy carriers and injects high-energy ones, and then letting the system evolve unitarily~\cite{supplementary}.

In Fig.~\ref{fig:DMFT_confronto} (a-b), we show that DMFT is indeed able to capture thermalization after the nonequilibrium excitation. To this extent, we monitor the distribution function $ F( \omega, t ) = \Im G^< (\omega, t)/2 \pi \mathcal{A} (\omega,t)$, where $G^<$ is the lesser component of the local Green's function (suppressing spin index since we consider a paramagnetic state) and $\mathcal{A}(\omega,t)=-\frac{1}{\pi} \Im G^R(\omega+i0,t)$ the spectral function. 
The distribution function $F(\omega,t)$ evolves from the initial sharp, low-temperature shape (essentially a Fermi distribution) to a much smoother one, matching  the equilibrium distribution at the final temperature 
$\beta=1.31$ (dashed lines). The spectral function $\mathcal{A}(\omega,t)$ evolves accordingly and thermalizes to the final temperature (see dashed line).


Thermalization within non-equilibrium DMFT has been reported before~\cite{eckstein2009thermalization,Aoki2014,peronaci2018resonant}, however its origin and physical interpretation has not been discussed so far. In the following, we address the question of how DMFT is able to describe self-consistently the emergence of thermalization, wherein the system acts as its own bath. 
As we are going to show, thermalisation can only arise because of DMFT self-consistency and it implies the development of a sharp thermalisation front whose properties we will discuss in detail.

 
\emph{Emergence of a Thermalization Front  -  } 
To understand the emergence of thermalization within DMFT we will take a radically different perspective and look at the problem from the point of view of the impurity, as its quantum bath is self-consistently updated. This approach to DMFT has been used to elucidate various features of equilibrium strongly correlated systems~\cite{Georges1996,florens2007nanoscale,held2013poor}, particularly the emergence of a Mott transition~\cite{georges1992numerical,rozenberg1992mott}.
Instead of making the DMFT converge at each time step, we first evolve the dynamics of the impurity in a given bath up to long times. Then, by imposing the self-consistency, we obtain the new bath for the next iteration $n$.
This amounts, in practice, in exchanging the order of limits between $t\rightarrow\infty$ (long-time limit) and $n\rightarrow\infty$ (DMFT self-consistency).

In Fig.~\ref{fig:DMFT_confronto} (c-d), we depict the dynamics of the double occupation $D_n(t)=\frac{1}{N_s}\sum_j \langle\hat{n}_{j\uparrow}\hat{n}_{j\downarrow}\rangle$ and the total energy $E_n(t)=\langle \hat{H}\rangle$, for different steps $n$ in the DMFT iteration, compared to the fully converged DMFT solution ($n=\infty$). After a sharp increase at short times, corresponding to the interval of photoexcitation (see red region),  the dynamics of $D_n(t)$
decays, reaching a plateau value (thick black line) corresponding to the fully converged DMFT equilibrium solution and ultimately going back at long times to the initial equilibrium value $D_{n=0}(t=0)$ (grey line with arrows). This long-time limit is fixed by the initial value independently on the iteration $n$ since the bath, at long enough times, is always in equilibrium at the initial temperature. As we increase $n$ and iterate the DMFT loop we see that the time scale $\tau^*_n$, separating the crossover between final thermal equilibrium and initial condition, grows with $n$, i.e. the dynamics spends a significant amount of time along the fully converged DMFT equilibrium solution (thick black line) before ultimately going back to the initial state.  A similar result is evident for the total energy $E_n(t)$ (panel d): After the excitation, the energy remains constant up to times $\tau^*_n$ but ultimately returns to the initial value, since the impurity exchanges energy with the non-equilibrium bath and thermalizes back to the initial state.  For $n\rightarrow\infty$,  the energy is 
strictly conserved, i.e., it remains constant after the end of the excitation,
as expected for a fully self-consistent DMFT solution with SPT impurity solver. 
In Fig.~\ref{fig:DMFT_confronto} (e-f), we depict the same quantities in the $n,t$ plane to track the evolution of the time scale $\tau^*_n$ at which the system first reaches the final equilibrium state. We observe the emergence of a sharp \emph{thermalisation front}  moving ballistically with $n$, $\tau^*_n\sim n/v$, separating the initial state fixed point (for $t>n/v$) from the final thermal one  (for $t<n/v$). If we fix the iteration $n$ and send the time $t\rightarrow\infty$, the bath will ultimately bring the system back to its initial value, and the memory of initial condition is never lost. On the contrary, at fixed $t$ and $n\rightarrow\infty$ we see that the system thermalizes.

\begin{figure*}
 	\centering \includegraphics[width=\mygraphicwidth]{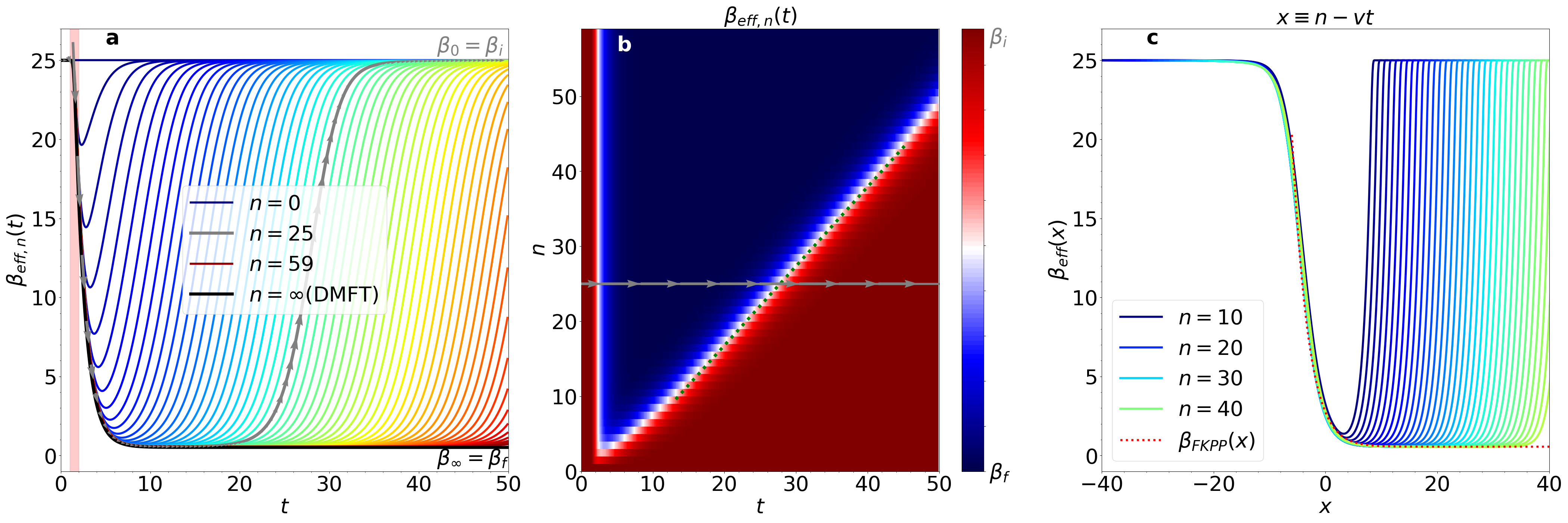}
	\caption{ {\bf Thermalization Front for the Effective Temperature in SPT}: Panel \textbf{a} describes the dynamics of the effective temperature for each iteration $n$, compared to the fully self-consistent DMFT solution (black line) as obtained with second-order self-consistent perturbation theory (SPT) and QBE. $\beta_{n}(t)$ is obtained in both cases from the derivative at zero frequency of the distribution function (see main text). We see clearly that the effective temperature first increases and then goes back to the initial value. This allows us to define a sharp front moving ballistically. 
    In Panel \textbf{b}, dashed green line, we plot the velocity of the front. 
    In Panel \textbf{c}, we rescale the data on the effective temperature and demonstrate a travelling wave solution with front velocity $v=1.06$ obtained from panel \textbf{b}. Red dotted curve: $\beta_{\text{FKPP}}$ is the asymptotic form of the front obtained from the analysis of the FKPP equation, as discussed in detail in Ref.~\cite{supplementary}.
    }
	\label{fig:DMFT_front}
\end{figure*}
The emergence of this front is a direct consequence of the system \emph{trying to act as a bath for itself}, as implemented by the DMFT self-consistency on the Anderson impurity model.  Indeed, the initial bath in equilibrium at temperature $\beta_i$ makes the local interacting impurity thermalize at its own temperature at long-times via the Kondo effect, thus never truly loosing memory of the initial condition. However, the impurity is locally excited by the non-equilibrium perturbation and its Green's function $G_{\sigma}(t,t')$  is out of equilibrium (breaking time-translation-invariance) up to a certain time scale $\tau_n$, before going back to the initial equilibrium. At the next step of the iteration, the DMFT bath is therefore highly-excited and out of equilibrium, roughly up to the same time scale $\tau_n$. As a result, the impurity dynamics will respond to this excited bath with some delay, leading to a relaxation time scale $\tau_{n+1}>\tau_n$ (i.e. a front) at which the impurity will locally heat up, through a mechanism akin to non-equilibrium induced decoherence~\cite{rosch2001kondo,kehrein2005scaling,mitra2006nonequilibrium,schiro2014transient}, before relaxing back to the initial equilibrium. It is therefore the transient out-of-equilibrium nature of the bath, together with the local interaction at the impurity site, that allows the emergence of heating and thermalisation in DMFT, ultimately converting the initial bath distribution function at $\beta_i$ into the final equilibrium one at a higher temperature $\beta_f$. We expect such a thermalization front to remain sharply defined even in the strongly correlated Fermi-Liquid regime or Mott insulating phase at finite temperature, since the interacting quantum impurity would eventually thermalize back to its own bath. The time scales for this front to form might strongly depend on the value of the interaction,as we can verify in the weak coupling regime~\cite{supplementary}.

%

\emph{ Thermalization front for Effective Temperature and FKPP Equation --}
To gain further insights on the thermalization front in the DMFT dynamics of the Hubbard model, it is useful to reduce the complexity and the richness of the information contained in the full DMFT solution and focus on a coarse grain description of the dynamics. We employ to this extent the recently developed Quantum Boltzmann Equation (QBE) framework for DMFT, which only assumes a separation
of timescales, taken into account through the gradient approximation for convolutions in time~\cite{Picano2021}.
The starting point is the QB equation for the distribution function $F( \omega, t )$
\begin{align}
	\label{QB_eq}
	\partial_t F ( \omega, t ) = I [F(\omega,t) ]
\end{align}
where $I[F]$ is the scattering integral, 
$I_{\omega} [F] = -i \{\Sigma^<(\omega,t)+F(\omega,t) [\Sigma^R(\omega,t)-\Sigma^A(\omega,t)] \}$,
taking contribution from both interaction, DMFT bath and excitation protocol~\cite{supplementary}.
The main advantage of the QBE is that it allows for a much longer time propagation~\cite{Picano2021} and
a characterization of the dynamics in terms of physically transparent quantities.
It enables the definition of a time-dependent effective temperature $T_{\rm eff}(t)=1/\beta_{\rm eff}(t)$ extracted from the low-frequency component of the distribution function $\beta_{\rm eff}(t) \equiv -4 \partial_\omega  F (\omega, t)|_{\omega=0}$.
We can use the effective temperature to characterize the build-up of a sharp thermalisation front. Specifically, we introduce an effective temperature, $\beta_{\rm eff, n}(t)=-4 \partial_\omega  F_n (\omega, t)|_{\omega=0}$, for each DMFT iteration $n$ that we plot in Fig.~\ref{fig:DMFT_front} (a-b). We observe a clear travelling front that separates the short time dynamics, at which the inverse temperature quickly reaches the value $\beta_f$, from the long-time dynamics, where the memory of initial condition is restored ($\beta_i$). This front travels ballistically and brings away the initial condition dependence, leaving the system in a well defined thermal state. We can confirm the travelling wave solution by collapsing the dynamics of the effective temperature as $\beta_n(t)\equiv \beta(x=n-vt)$ with the velocity extracted from the thermalization front. In Fig.~\ref{fig:DMFT_front} (c), we observe the nice collapse into a stable front, reminiscent of a traveling wave equation. 

We can derive explicitly an equation for the deviation from equilibrium of the time-dependent effective temperature.  It is an FKPP equation, which therefore provides a natural framework to study the thermalization front.  More precisely, we parametrize the distribution function at iteration $n$ as $F_n(\omega,t)=f_{\beta_f}(\omega) +u_n(t) [f_{\beta_i}(\omega)-f_{\beta_f}(\omega)]$ where $f_{\beta}(\omega)$ is the equilibrium Fermi distribution function and $u_n(t)$ parametrizes the deviations from the two equilibrium fixed points ($u_n\equiv0,1$ corresponding respectively to $\beta_{f,i}$). If we now plug this parametrization into Eq.~(\ref{QB_eq}), and evaluate the different contributions to the scattering integral as a function of $u_n(t)$, we obtain a discrete FKPP equation with quartic non-linearity~\cite{brunetderrida2015}
\begin{align}\label{eqn:FKPP}
\Dot{u}_n(t) = C [u_{n-1}(t)-u_n(t)]+ \Phi(u_n) +\eta_n(t) 
\end{align}
where $\eta_n (t) \equiv A(t) +B(t) u_n$ is the driving field and $\Phi (u_n) \equiv E u_n+F u_n^2+G u_n^3$ is the non-linear functional that comes from the electronic self-energy. We emphasize that the mapping to FKPP relies only on the self-consistent nature of the DMFT bath, allowing to relate the distribution function at step $n+1$ with the one at step $n$ and on the fact that the self-energy is a functional of the impurity Green's function, and not on the specific choice of impurity solver (here QBE/SPT). A different choice would result in a different non-linear functional $\Phi(u_n)$.
The coefficients $A,\ldots,G$ in Eq.~(\ref{eqn:FKPP}) are all known in terms of the spectral function of the equilibrium state~\cite{supplementary}.
Solving for $u_n(t)$, we obtain a sharp front propagating ballistically which matches very well the full numerical solution of the QBE~\cite{supplementary}. 
In particular, we show in panel (c) the comparison between the rescaled QBE solution and the asymptotic form of the front $\beta_{\text{FKPP}}$ obtained from the long-time analysis of the FKPP equation~\cite{supplementary}, showing perfect agreement.

\emph{Discussion -- }In this work, we discussed how quantum thermalisation emerges for interacting fermions in infinite dimension. This limit, as described by non-equilibrium DMFT, naturally allows  to understand the emergence of a self-consistent, quantum bath for the system itself. We have shown that thermalisation arises as a sharp travelling front, which separates the long-time thermal equilibrium fixed point from the initial condition. Remarkably, we have demonstrated that the properties of the front are described by a travelling-wave equation of FKPP form.  It is tempting to interpret the DMFT iteration $n$ as a fictitious space coordinate which accounts for the distance between the boundary, where the non-equilibrium bath is imposed, and the center~\cite{florens2007nanoscale}. In this picture, the thermalisation front that we have discussed could be seen as the result of the information propagation from the boundary. In this respect, the ballistic front is consistent with the Lieb-Robinson bound and with the non-interacting nature of the bath in infinite dimensions. An intriguing question is how this picture would be modified in finite dimensions and whether a diffusive front could emerge. Cluster DMFT extensions to finite dimensions together with finite-size numerical simulations  could be explored to investigate this issue. 
While we have focused here on fermions, we expect our thermalization front picture to hold also for the large connectivity limit of bosons~\cite{strand2015nonequilibrium,scarlatella2021dynamical,shih2022anharmonic}.
The result obtained here provides a new lens through which analyze and characterise thermalising and non thermalising dynamics. A natural avenue for future applications is the case of disordered quantum many-body systems, using statistical DMFT~\cite{Miranda_2005}. We can expect that in an Anderson localised phase, the ballistic front would give way to a localised one, which never looses memory of the initial condition. In presence of disorder and interactions, we 
 argue a competition between thermalizing fronts and pinning due to disorder, which could shed new light on the stability of many-body localisation.

\emph{Acknowledgements - }We acknowledge financial support from the ERC consolidator grant No.~101002955 - CONQUER. We acknowledge computational resources on the Coll\`ege de France IPH cluster.

\clearpage
\appendix
\widetext
\begin{center}
	\textbf{\large \centering Supplemental Material: Quantum Thermalization via Travelling Waves}
\end{center}
\vspace{1cm}
\setcounter{equation}{0}
\setcounter{figure}{0}
\setcounter{table}{0}
\setcounter{page}{1}

In this Supplemental Material, we discuss: 
\begin{enumerate}
	\item The implementation of the fully self-consistent non-equilibrium DMFT for the photoexcited Hubbard model with SPT solver and the Quantum Boltzmann approach.
	\item The DMFT and QBE algorithm implemented as a step-by-step procedure to follow the development of thermal equilibrium.
	\item Additional Numerical Results: Dependence on the interaction and comparison between QBE and full DMFT.
	\item The derivation of the FKPP equation, the discussion of its properties and comparison with DMFT results.
\end{enumerate}

\section*{Nonequilibrium DMFT solution of the Hubbard Model}
\label{sec::full_DMFT_solution}
Within DMFT~\cite{Georges1996}, one maps the lattice model \eqref{ham:hubb} onto an effective single-site impurity model. The coupling of the latter to the environment is described by the hybridization function $\Delta(t,t')$, which is self-consistently determined such that the local ($\bm k$-averaged) lattice Green's function, $G_{\text{loc}}(t,t')$, coincides with the impurity Green's function, $G_{\text{imp}}(t,t')$. 
In DMFT, one assumes the lattice self-energy to be local in space (independent of $\bm k$) and requires it to be identical to the impurity self-energy: $\Sigma_{\text{loc}}(t,t') \equiv \Sigma_{\text{imp}}(t,t')$ . In detail, the impurity model is defined by an action 
\begin{align}
	\mathcal{S}
	=
	-i\int_\mathcal{C} dt\, H_{\text{loc}}(t)-i
	\int_\mathcal{C} dt dt' 
	\sum_\sigma
	c_\sigma^\dagger(t) \Delta_\sigma(t,t')c_\sigma(t'),
\end{align}
in terms of the self-consistent hybridization function $\Delta(t,t')$ and the time-local energy $H_{\text{loc}}(t)$.
The latter is expressed in terms of the chemical potential $\mu$ and $h_\sigma(t)$, the single-particle Hamiltonian in the impurity model: $H_{\text{loc}}(t)=h(t)-\mu$.
The single-particle Hamiltonian $h(t)$ can contain, e.g., the Hartree self-energy; in this case, $h(t)=U n(t)$ with the density $n= \sum_\sigma n_{\sigma}$. Since we restrict ourselves to the half-filled system ($\mu=U/2$ and $n_\sigma=\frac{1}{2}$), for us it is always $H_{\text{loc}}(t)=0$.
In the following, we will omit the $\sigma$ index for simplicity.

The non-interacting impurity Green's function $\mathcal{G}$ is determined by the Dyson equation
\begin{align}
	\label{Gweiss}
	\mathcal{G}^{-1}(t,t')=[i\partial_t -H_{\text{loc}}(t)]\delta_{\mathcal{C}}(t,t') -\Delta(t,t')
\end{align}
The interacting impurity Green's function is given by
\begin{align}
	\label{Dimp2}
	G_{\text{imp}}^{-1}(t,t')=\mathcal{G}^{-1}(t,t')-\Sigma_{\text{int}}(t,t')
\end{align}
where 
\begin{align}
	\label{Sigma_int}
	\Sigma_{\text{int}}(t,t')
	=
	\Sigma_{U}(t,t')+\Gamma(t,t')
\end{align}
is the impurity self-energy.  $\Sigma_{\text{int}}$ is given by two contributions: The Hubbard interaction ($\Sigma_{U}$) and the self-energy $\Gamma$ which describes the coupling with a non-equilibrium bath that mimics photo-excitation. The former is self-consistently determined using the Self-Consistent second-order Perturbation Theory impurity solver (SPT), as described in Eq.~\eqref{SPT}, while the latter is defined in Eq.~\eqref{Gamma}.
For the semi-elliptic density of states considered in the main text, the  DMFT self-consistency can be formulated in closed form, and the hybridization of the impurity model is simply given by \cite{Georges1996,Aoki2014}
\begin{align}
	\label{bethe}
	\Delta(t,t')=t_h^2G(t,t')
\end{align}
in terms of the local interacting Green's function $G$.
To summarize, for each $(t,t')$ on the Keldysh contour $\mathcal{C}$~\cite{Aoki2014}, self-consistency is reached by iterating till convergence the following loop:
\begin{equation}
\begin{cases}
\Delta(t,t')&=t_h^2G(t,t')  \nonumber \\
\mathcal{G}^{-1}(t,t')&=[i\partial_t +\mu -h(t)]\delta_{\mathcal{C}}(t,t') -\Delta(t,t') \nonumber \\
\Sigma_{\text{int}}(t,t')&=\Sigma_{U}(t,t')+\Gamma(t,t') \nonumber \\
G^{-1}(t,t')&=\mathcal{G}^{-1}(t,t')- \Sigma_{\text{int}}(t,t')       
\end{cases}
\label{DMFT}
\end{equation}
The above equations are solved using the NESSi simulation package~\cite{Schuler2020}. The total energy of the system is given by:
\begin{align}
	\label{energyfull}
	E(t)
	= -2i(\Delta*G)^<(t,t)- i (\Sigma_{\text{int}}*G)^<(t,t)
\end{align}
The first and second term represent the kinetic and interaction energy, respectively, with a factor two in the kinetic energy for the summation over spin components.
The double occupancy is given by~\cite{Eckstein2010}:
\begin{align}
	D(t)=- i (\Sigma_{\text{int}}*G)^<(t,t)/U(t)+\frac{1}{2} n(t).
\end{align}

\subsection{SPT self-energy} \label{sec:SPT}

The SPT approximation is  based on a second-order diagrammatic evaluation of the self-energy in terms of the fully interacting impurity Green's functon $G$. The associated self-energy reads
\begin{align}
	\Sigma_U(t,t')
	=
	U(t)U(t')
	G(t,t')
	G(t,t')
	G(t',t)
	\label{SPT}
\end{align}
SPT is a conserving approximation, since the self-energy is written as a functional of the fully dressed Green's function. It is therefore particularly suited to study questions related to thermalization and energy transfer between impurity and bath. It is however limited to weak coupling~\cite{Aoki2014},
we therefore restrict ourselves in the present study to $U=2$, so that the expansion of the self-energy $\Sigma$ in terms of the Green's function $G$ is appropriate both at equilibrium and in non-equilibrium. 
%
%
\subsection*{Excitation protocol} 
\label{sec:ex_prot}
In order to simulate a photo-doping excitation, the system is shortly coupled with a fermionic bath with density of states
\begin{align}
	\label{Abath}
	\mathcal{A}_{{\text{bath}}}(\omega)= \mathcal{A}_b (\omega-\omega_0) + \mathcal{A}_b (\omega+\omega_0)
\end{align}
consisting of two smooth bands with bandwidth $W_{\rm bath} = 4$ centred around the energies $\omega_0 = \pm 2.5$~\cite{Picano2023_inhomogeneous}. We choose
\begin{align}
	\mathcal{A}_b (\omega) = \frac{1}{\pi} \cos^2 (\pi \omega /W_{\text{bath}})
\end{align} 
in the interval $[\omega_0-W_{\rm bath}/2,\omega_0+W_{\rm bath}/2]$.
The occupied and unoccupied density of states have spectral shapes given by $\mathcal{A}_{\rm bath}^{<} (\omega)= \mathcal{A}(\omega-\omega_0)$ and $\mathcal{A}_{\rm bath}^{>} (\omega)= \mathcal{A} (\omega+\omega_0)$, respectively. 
During the whole time-evolution of the system, the bath occupation $f_{\text{bath}}(\omega)=f_{-\beta_i}(\omega)$ is taken to be fixed at negative temperature Fermi-Dirac distribution (population inversion). In this way, the coupling of the system with the bath mimics photo-excitation.
$\beta_i$ corresponds to the initial (inverse) temperature of the system (in our calculations, $\beta_i \equiv \frac{1}{T_i}=25$).

The fermionic bath adds a local contribution to the electronic self-energy $\Sigma_{\text{int}}$ in Eq.~(\ref{Sigma_int}), given by:
\begin{align}
	\label{sphonto}
	\Gamma(t,t')=V(t)G_{\text{bath}}(t,t')V(t')^*,
\end{align}
with time-dependent profile 
\begin{align}
	\label{V}
	V(t) = V_0 \sin^2 (\pi t/T_0) \theta (t) \theta (t_0-t),
\end{align}
where $V_0=1$, $T_0=1$ and $t_0 =1$ in units of hopping times (see main text). 
The bath Green's function $G_{\text{bath}}(t,t')$ is defined as
\begin{align}
	G_{\text{bath}}^R(t,t')
	&=
	-i
	\theta(t-t')
	\int d\omega \,e^{-i\omega(t-t')}\mathcal{A}_{\text{bath}}(\omega),
	\\
	G_{\text{bath}}^<(t,t')
	&=
	i
	\int d\omega\,
	e^{-i\omega(t-t')} f_{\text{bath}}(\omega)\mathcal{A}_{\text{bath}}(\omega)
\end{align}
on the Keldysh contour $\mathcal{C}$~\cite{Aoki2014}.

The coupling with the fermionic bath results in two bumps in the distribution function for short times after the excitation: A \textit{positive} bump around $\omega_0 = +2.5$ with bandwidth $W_{\rm bath} = 4$, which corresponds to the \textit{injection} of electrons above the Fermi energy, and a \textit{negative} bump around $\omega_0 = -2.5$ with bandwidth $W_{\rm bath} = 4$, which corresponds to the \textit{depletion} of electrons below the Fermi energy.
	In our simulations, the coupling with the fermionic bath lasts from $t=1$ to $t=2$ (vertical red shaded area in Fig.~1 (c)-(d)) and, indeed, for $t=2$, we see the appearance of the two bumps. They are visible till $t=2.8$ and then, as thermalization develops, they start to disappear. In the end, the system approaches a new equilibrium state with $T_f >T_i$ ($\beta_f = 1.31 < \beta_i =25 $ in Fig.~(1) of the main text, for $U=2$, and $\beta_f = 1.63 < \beta_i =25 $ in Fig.~\ref{fig:DMFT_confronto_U_1} of the SM, for $U=1$).

\subsection{QBE formulation of the problem}
\label{sec:QBE}

The time evolution of the system in Eq.~\eqref{ham:hubb} can be computed by means of a non-perturbative QBE for the energy distribution function
\begin{align}
	\label{FG}
	F( \omega, t ) = \frac{\Im G^< (\omega, t)}{2 \pi \mathcal{A} (\omega,t)}
\end{align} 
solved in the framework of DMFT~\cite{Picano2021}, where $G^<$ is the lesser component of the interacting Green's function $G$ and $\mathcal{A}(\omega)=-\frac{1}{\pi} \Im G^R(\omega+i0)$ the spectrum. 
The QBE gives an equation 
\begin{align}
	\label{QBE}
	\partial_t F ( \omega, t ) = I_\omega[F(\omega,t), \cdot ]
\end{align}
for the evolution of the distribution, with scattering integral:
\begin{align} 
	\label{scatt_int}
	I_{\omega} [F] = &-i \{\Sigma^<(\omega,t)+\Sigma^R(\omega,t)F(\omega,t)-F(\omega,t) \Sigma^A(\omega,t) \}
\end{align}
where $\Sigma(\omega, t)=\Sigma_{\text{int}}(\omega, t)+\Delta(\omega, t)$.
The evaluation of the functionals $\Sigma_{\text{int}}(\omega,t)$ and $\mathcal{A}(\omega,t)$ at each time $t$ is done iteratively by solving a non-equilibrium steady-state (NESS) impurity problem with a DMFT impurity solver of choice (SPT in our case). 
In practice, for each time $t$, $F(\omega, t+h)\equiv \bar F(\omega)$ is given by the QBE Eq.~\eqref{QBE} and 
is not updated during the NESS loop described below, while
$\Sigma_{\text{int}}(\omega,t+h) \equiv \bar \Sigma_{\text{int}}(\omega)$ 
and $\mathcal{A}(\omega,t+h) \equiv \bar{\mathcal{A}}(\omega)$ are calculated self-consistently in the following way:
\begin{enumerate}
	\label{NESSloop}
	\item 
	Start from a guess for $\bar \Sigma_{\text{int}}(\omega)$ (if you are at the very first timestep, $t=0$, otherwise start from $\Sigma_{\text{int}}(\omega,t)$ calculated at the previous timestep) and solve the steady-state equation for $\bar G^R(\omega)$, 
	\begin{align}
		\label{G_R}
		& \bar G^R(\omega)
		= 
		[\omega +i0^+ -\bar H_{\text{loc}}(t) - \Delta^R(\omega)- \bar \Sigma_{\text{int}}^R(\omega)]^{-1}.
	\end{align}
	in order to determine $ \bar{ \mathcal{A}}(\omega)=-\frac{1}{\pi} \Im \bar G^R(\omega+i0)$.
	We recall that in our case (half-filling) it is always $\bar H_{\text{loc}}(t) = \bar h(t) - \mu =0$. 
	\item
	Determine the lesser Green's function from the given distribution function, using the steady-state variant of the fluctuation-dissipation theorem:
	\begin{align}
		\bar G^<(\omega) = 2\pi i \bar F(\omega) \bar{\mathcal{A}}(\omega)	
	\end{align} 
	In NESS, $\bar F(\omega)$ does not need to be a Fermi-Dirac distribution and, indeed, in general it is not.
	\item
	Use the self-consistency Eq.~\eqref{bethe} to fix the hybridization function of the effective steady state impurity model,
	\begin{align}
		\label{Delta_QBE}
		\Delta^{R,<}(\omega)=t_h^2\bar G^{R,<}(\omega)
	\end{align}
	\item
	Solve the impurity model, by means of the SPT impurity solver.
	In the time-translational invariant case (NESS loop), the lesser and retarded components of the electron-electron interaction  $\Sigma_{U}(t)=U(t)^2 G(t)G(-t)G(t)$ read:
	\begin{align}
		\label{Sigma_U}
		\Sigma_{U}^{<(>)}(t) &=U^2(t)[G(t)G(-t)]^{<(>)}G^{<(>)}(t) \nonumber \\
		&=U^2(t)G^{<(>)}(t)G^{>(<)}(-t)G^{<(>)}(t) \nonumber \\
		&=U^2(t)G^{<(>)}(t)[-G^{>(<)}(t)]^\dagger G^{<(>)}(t) \\
		\Sigma_{U}^R(t) &= \Sigma_{U}^>(t) - \Sigma_{U}^<(t)+ \Sigma_{U}^A(t) \nonumber \\
		&= U^2(t)G^>(t)[-G^<(t)]^\dagger G^>(t)- \Sigma_{U}^<(t), \,\,\,\, \forall t>0.
		\label{Sigma_U_ret_ss}
	\end{align} 
	In practice, start from $G(\omega)$, anti-Fourier transform it in $G(t)$, get $\Sigma(t)$ from Eqs.~\eqref{Sigma_U}-\eqref{Sigma_U_ret_ss}, and Fourier transform it to get $\Sigma_U(\omega)$.
	\item
	Set $\bar \Sigma_{\text{int}}(\omega)=\Sigma_{U}(\omega)+\Gamma(\omega)$, and iterate step 1) to 5) until convergence.
	The fermionic bath $\Gamma$ in Eq.~\eqref{sphonto}, in the steady-state formalism becomes:
	\begin{align}
		\label{Gamma}
		\Gamma^{<(>)} (\omega,t) &=(-)2\pi i V^2(t) \mathcal{A}_{\text{bath}}^{<(>)} (\omega), \\
		\Gamma^R(\omega,t) &= - i \pi V^2(t) \mathcal{A}_{\text{bath}}(\omega)
		\label{Gamma_ret}
	\end{align}
	with time-dependent profile coupling $V(t)$ defined in Eq.~\eqref{V}.
	%
	
\end{enumerate}
The DMFT self-consistency serves as a way to evaluate $\Sigma^{\text{NESS}}[F_G(\cdot,t)]$ (as well as $\mathcal{A}^{\text{NESS}}[F_G(\cdot,t)]$ and $\Delta^{\text{NESS}}[F_G(\cdot,t)]$). 
The differential equation \eqref{QBE} is  then solved  using a Runge-Kutta algorithm at fourth order and, once one gets $F(\omega, t+h)$,  the new NESS loop at time $t+h$ is solved.  

The  total energy of the system at (average) time $t$ is
\begin{align}
	E(t)=\frac{1}{2 \pi}\int d\omega \left \{-2i [\Delta(\omega,t)G(\omega,t)]^<- i [\Sigma_{\text{int}}(\omega,t)G(\omega,t)]^< \right \}
	\label{energy_QBE}
\end{align}
while the double occupancy is given by:
\begin{align}
	D(t)=\frac{1}{2 \pi} \int d\omega \left \{- i [\Sigma_{\text{int}}(\omega,t)G(\omega,t)]^< \right \} + \frac{1}{2} n(t).
\end{align}
The non-perturbative QBE approach, which relies on some assumptions (the main of which is the separation of timescales, see~\cite{Picano2021}), gives a computational effort which scales like $\mathcal{O}(t_{\text{max}})$, as compared to the $\mathcal{O}(t_{\text{max}}^3)$ of the Noneq-DMFT solution. 
Moreover, in QBE the memory occupied does not depend on $t_{\text{max}}$, while in Noneq-DMFT it scales like  $\mathcal{O}(t_{\text{max}}^2)$.

\section*{Noneq-DMFT solution: step-by-step approach}
\label{sec::full_DMFT_solution_alt}

In the alternative formulation of the Noneq-DMFT problem, we start from the equilibrium solution of the DMFT system of equations at $T=T_i$, $U=U_i$. For the very first DMFT iteration, $n=0$, the photo-excitation bath is switched off for all times: $\Gamma_0(t,t')\equiv0$.
Having reached self-consistency at equilibrium, we propagate the solution by means of the usual Dyson's equations~\cite{Aoki2014} for all $(t,t')$.
Here we reach DMFT-convergence for each $(t,t')$, as we did in the normal approach, the only difference being the fact that the excitation bath is always off, so the system is in equilibrium during the whole dynamics.
Actually, we are imposing that $\Delta_0(t,t')$, $G_0(t,t')$, $\Sigma_{\text{int},0}(t,t')$ and $\mathcal{G}_0(t,t')$ are time translational invariant (TTI) functions in all their components.

For DMFT-iterations $n=1,2,\dots, n_{\text{max}}$, we switch on the excitation bath $\Gamma$ and propagate in time up to $t_{\rm max}$ without imposing the DMFT self-consistency. Actually, for each $(t,t')$ we solve only once the system of equations:
\begin{equation}
\begin{cases}
\Delta_n(t,t')&=t_h^2G_{n-1}(t,t')  \nonumber \\
\mathcal{G}_n^{-1}(t,t')&=[i\partial_t +\mu -h(t)]\delta_{\mathcal{C}}(t,t') -\Delta_n(t,t') \nonumber \\
\Sigma_{U,n}(t,t')&=U(t)U(t') G_n(t,t') G_n(t,t') G_n(t',t) \nonumber \\
\Gamma(t,t')&=V(t)G_{\text{bath}}(t,t')V(t')^* \nonumber \\
\Sigma_{\text{int},n}(t,t')&=\Sigma_{U,n}(t,t')+\Gamma(t,t') \nonumber \\
G^{-1}_n(t,t')&=\mathcal{G}_n^{-1}(t,t')- \Sigma_{\text{int},n}(t,t') 
\end{cases}
\end{equation}
We observe that the hybridization function $\Delta_n(t,t')$ depends on the $G_{n-1}$ at the previous DMFT-iteration.
The crucial property of this bath, as already mentioned in the main text, is that it is explicitly non-TTI, at least for early times after the excitation; in the long-time limit, it will approach the intial temperature equilibrium $\Delta_n^{ss}$. This is due to the fact that $G_{n-1}$ keeps the memory of the initial temperature $T_i$.

We impose that convergence is reached for $n=n_{\text{max}}$ such that 
\begin{align}
	\sum_{t,t'} \abs{G_{n_{\text{max}}}(t,t')-G_{n_{\text{max}}-1}(t,t')}< 10^{-6}
\end{align}
where $10^{-6}$ is the error threshold that we arbitrarily fix.
The total energy of the system is given by:
\begin{align}
	E_n(t)
	= -2i(\Delta_n*G_n)^<(t,t)- i (\Sigma_{\text{int},n}*G_n)^<(t,t)
\end{align}
with a factor two in the kinetic energy due to the summation over the two spin components. The double occupancy is given by:
\begin{align}
	D_n(t)=- i (\Sigma_{\text{int},n}*G_n)^<(t,t)/U(t)+\frac{1}{2} n_n(t).
\end{align}

\subsection{QBE: step-by-step approach}
\label{sec:QBE_alternative}
In the method considered so far, for each timestep, given the distribution function $F(\omega,t+h)$, which comes from the QBE in Eq.~\eqref{QBE}, the spectrum $\mathcal{A}$, the hybridization function $\Delta$ and the self-energy $\Sigma$ are updated by reaching DMFT self-consistency in the NESS loop. 
Only after convergence has been reached, the distribution function  $F(\omega,t+2h)$ is calculated and the new NESS loop restarts. 

Here, we consider the case in which DMFT convergence is reached as a function of time rather than for each single timestep. 
At the first iteration, $n=0$, and the system is in equilibrium at the initial temperature $T_i$.
The equilibrium functions $G_{n=0}(\omega,t=0)$, $\Delta_{n=0}(\omega,t=0)$ and $\Sigma_{n=0}(\omega,t=0)$ are determined from the same NESS loop described in the previous Section in step 1) - 5), and are called $G_{0}(\omega)$, $\Delta_{0}(\omega)$ and $\Sigma_{0}(\omega)$.  
When ($n=0$), the excitation bath $\Gamma$ is set to zero for all the timesteps, and the equilibrium Green's functions are not updated in time: 
\begin{equation}
\begin{cases}
G_{n=0}(\omega,t) \equiv G_0(\omega) ,\,\,\,\, \forall t \nonumber \\
\Delta_{n=0}(\omega,t)  \equiv \Delta_0(\omega) ,\,\,\,\, \forall t \nonumber \\
\Sigma_{n=0}(\omega,t)  \equiv  \Sigma_0(\omega) ,\,\,\,\, \forall t
\end{cases}
\end{equation}
Starting from these functions, several DMFT iterations $n$, each of which goes from $t=0$ to $t=t_{\text{max}}$, are performed until DMFT convergence is reached. Each full time propagation $t \in [0,t_{\text{max}}]$ is identified by the index $n$ ($n \geq 1$). 
They all start from the same initial condition at $t=0$:
\begin{equation}
\begin{cases}
G_n(\omega,t=0)=G_0(\omega) ,\,\,\,\, \forall n \geq 1 \nonumber \\
\Delta_n(\omega,t=0)=\Delta_0(\omega) ,\,\,\,\, \forall n \geq 1 \nonumber \\
\Sigma_n(\omega,t=0)=\Sigma_0(\omega) ,\,\,\,\, \forall n \geq 1
\end{cases}
\end{equation}
Given this initial condition, the DMFT loop for $t \in [0,t_{\text{max}}]$, for each full time propagation $n \geq 1$, looks like:
\begin{enumerate}
	\label{NESSloop1}
	\item
	Use the self-consistency Eq.~\eqref{bethe} to fix the hybridization function of the effective steady state impurity model:
	\begin{align}
		\begin{cases}
			\label{Deltan1}
			\Delta_1^{R,<}(\omega,t)=t_h^2 G^{R,<}_0(\omega,t=0) , \,\,\,\ \text{for} \,\,\,\ n=1 \\
			\Delta_n^{R,<}(\omega,t)=t_h^2 G^{R,<}_{n-1}(\omega,t), \,\,\,\ \forall  n>1
		\end{cases}
	\end{align}
	We notice that, for $n=1$, $\Delta_1$ is proportional to $G_0$ for each time-step $t$, i.e., the impurity sees a self-consistent bath that is fixed at the initial temperature $T_i$ during the whole time propagation. Starting from $n>1$, for each timestep $t$, the impurity sees a bath which depends on the Green's function at the previous DMFT iteration $n-1$.
	\item 
	Update the retarded Green's function:
	\begin{align}
		& G_n^R(\omega,t)
		= 
		[\omega -H_{\text{loc},n}(t) - \Delta_n^R(\omega,t)- \Sigma_{\text{int},n}^R(\omega,t)]^{-1} ,\,\,\,\, \forall n \geq 1
	\end{align}
	and determine $  \mathcal{A}_n(\omega,t)=-\frac{1}{\pi} \Im  G^R_n(\omega+i0,t)$.
	We recall that in our case (half-filling) it is always $H_{\text{loc},n}(t) = h_n(t) - \mu =0$. 
	\item
	Determine the lesser Green's function from the given distribution function,
	\begin{align}
		G^<_n(\omega,t) = 2\pi i F_n(\omega,t) {\mathcal{A}}_n(\omega,t) ,\,\,\,\, \forall n \geq 1	
	\end{align} 
	\item
	Solve the impurity model, by means of the SPT as an impurity solver. The equations to determine $\Sigma_{U,n}(\omega,t)$ are the $n-$dependent version of Eqs.~\eqref{Sigma_U}-~\eqref{Sigma_U_ret_ss}.
	\item
	Set $\Sigma_{\text{int},n}(\omega,t)=\Sigma_{U,n}(\omega,t)+\Gamma(\omega,t)$. $\Gamma$ is given by the same Eqs.~\eqref{Gamma}-~\eqref{Gamma_ret} (there is no $n-$dependence in the quantities related to the bath).
	
	\item  
	Update $F_n$ by means of QBE Eq.~\eqref{QBE}:
	\begin{align}
		\partial_t F_n ( \omega, t ) = I [F_n(\omega,t), \cdot]
	\end{align}
	\item  Restart the loop step 1)- 6) for the next timestep $t+h$ till $t_{\text{max}}$ is reached. 
	Once $t_{\text{max}}$ is reached for the iteration $n$, the new iteration $n+1$ starts.
	Convergence is reached when, for each time $t$, the functions are not updated significantly from one iteration to the other, i.e., we require
	\begin{align}
		\sum_{\omega,t} \abs{ G_n(\omega,t)-G_{n-1}(\omega,t)} <10^{-6}
	\end{align}	  
\end{enumerate}
The  total energy of the system at iteration $n$ is
\begin{align}
	E_n(t)=\frac{1}{2 \pi}\int d\omega \left \{-2i [\Delta_n(\omega,t)G_n(\omega,t)]^<- i [\Sigma_{\text{int},n}(\omega,t)G_n(\omega,t)]^< \right \}
	\label{energy_QBE_n}
\end{align}
while the double occupancy is given by:
\begin{align}
	D_n(t)=\frac{1}{2 \pi} \int d\omega \left \{- i [\Sigma_{\text{int},n}(\omega,t)G_n(\omega,t)]^< \right \} + \frac{1}{2} n_n(t).
\end{align}

\section{Additional Numerical Results}

	\subsection{Nonequilibrium DMFT results for different values of U}
In this Section we present additional numerical results for the full nonequilibrium DMFT dynamics of the Hubbard model for different values of the interaction $U$, as compared to the ones for $U=2$, shown in the main text in Fig.~1. Specifically in panel (\textbf{a}) of Fig.~\ref{fig:DMFT_confronto_U_1} we plot the time evolution of the total energy after an energy excitation,for $U=1$ (initial inverse temperature $\beta_i=25$) and for different DMFT iteration number (coloured lines from blue to red as a function of the DMFT iteration number $n \in [0,99]$). The thermalization front is more clearly shown by plotting the data in the $(n,t)$ plane as we do in panel (\textbf{b}).
	We repeat this analysis for different values of interaction $U$ in the weak-to-intermediate coupling regime and extract the velocity of the thermalization front and its dependence from $U$, that we plot in panel  (\textbf{c}). We see that quite generally the velocity of the front decreases with interactions, as expected since the quasiparticle lifetime is reduced by entering the strongly correlated regime and so the time-scales for thermalization, roughly related to the inverse Kondo temperature of the associated impurity model. Still we can conclude that the picture of a thermalization front is robust to changes of the interaction within the weak to moderate coupling regime.


\begin{figure*}[h]
	\centering \includegraphics[width=\mygraphicwidth]
	{Figure01newU1.png}
		\caption{\label{fig:DMFT_confronto_U_1} {\bf Thermalization Front in Infinite Dimensions with DMFT for different values of $U$}: Time evolution of the total energy (\textbf{a}) after an energy excitation, for $U=1$ (initial inverse temperature $\beta_i=25$),  for different DMFT iteration number (coloured lines from blue to red as a function of the DMFT iteration number $n \in [0,99]$) and compared to the full self-consistent DMFT solution (black line). In panel  (\textbf{b}) same plot is shown in the $(n,t)$ plan showing clearly the ballistic front.	The evolution of the front velocity as a function of the interaction $U$ is plotted in panel  (\textbf{c}).}
\end{figure*}

\subsection{QBE thermalization dynamics}


	\begin{figure*}[t]
		\centering \includegraphics[width=\mygraphicwidth]
		{FigureQBEU2beta25h0pt01TVt1AmpVt0pt65tstpmax2000.png}
			\caption{\label{fig:QBE_confronto_U_2_Amp_pt65} {\bf Thermalization in Infinite Dimensions with DMFT plus QBE for $U=2$ and amplitude of the coupling with the fermionic bath $V_0=0.65$.}: Time evolution of the distribution function (\textbf{a}) and spectral function (\textbf{b}) after an energy excitation, for $U=2$ (initial inverse temperature $\beta_i=25$), leading to a thermal equilibrium fixed point. Full self-consistent QBE solution ($n=\infty$ in the notation of panels (\textbf{c})-(\textbf{d})).  {\bf Thermalization Front}: time evolution of the total energy (\textbf{c}) and double occupation (\textbf{d}) after an energy excitation, for different DMFT iteration number (coloured lines from blue to red as a function of the DMFT iteration number $n \in [0,50]$) and compared to the full self-consistent DMFT solution (black line). Emergence of a thermalization front in both energy and double occupation, panels (\textbf{e})-(\textbf{f})}.
	\end{figure*}

In this subsection we present the results of the  QBE+DMFT relaxation dynamics for each step $n$ of the DMFT self-consistency loop, to be compared with the equivalent analysis performed on the full DMFT data and shown in Fig.~1 of the main text. Specifically, in Fig.~\ref{fig:QBE_confronto_U_2_Amp_pt65}  we start from the same initial condition as in Fig.~1 of the main text ($U=2$ and $\beta_i=25$), and we apply the same photo-excitation, i.e. we couple the system with the same fermionic bath at negative temperature as in Fig.~1 for the same interval of time $\Delta t=1$. Concerning the amplitude of the coupling to the bath $V_0$, we choose it such that the final total energy the system approaches after the excitation for $n=\infty$ is the same as in the full-DMFT case ($E_\infty=-0.64$). Due to the non-perturbative nature of the QBE we employed, this corresponds also to the same final double occupancy ($D_\infty$), same final temperature $\beta_f=1.31$ and final spectral density $\mathcal{A}$.  We note that the value of $V_0=0.65$ used here is  slightly smaller as compared to $V_0=1$ used in Fig. 1 of the main text.  This can be explained by noticing that, due to the gradient approximation of the QBE, which fails for short instants of times after photoexcitation~\cite{Picano2021}, even if the excitation imposed to the system is the same, the systems actually absorbs a different amount of energy in the two cases. In panels (a-b) we present the dynamics of spectral function and distribution function in the fully self-consistent QBE approach, while panels (c-e) and (d-f) show respectively the evolution of the dynamics of double occupation and total energy step by step in the DMFT self-consistency. The appearance of the wave-front is evident in the QBE results like it was in the full-DMFT ones, and the velocity of the front is also comparable, confirming the agreement between the two approaches.

\section*{From Quantum Boltzmann to  FKPP Dynamics}

In this Section we derive the FKPP equation for the distribution function from the Quantum Boltzman/DMFT approach. We then discuss the key properties of the long-time solution of the FKPP equation and compare it with the DMFT results.

The study of the solutions of partial differential equations describing a moving interface from a stable to an unstable medium is a classical subject in mathematics, biology and physics. The prototype of such equations is indeed the FKPP equation~\cite{brunetderrida2015}
	\begin{equation}
	\frac{\partial u}{\partial t} = \frac{\partial^2 u}{\partial x^2} + f(u) \,\,,
	\end{equation}
	where the field \( u(x, t) \)  satisfies $0 \leq u(x,t) \leq 1$ and $f(u)$ is the nonlinear term.
	In our case, the FKPP equation describes the onset of thermalization in the photoexcited system;
	more in general, $u$ may represent the population density or concentration of a substance at position \( x \) and time \( t \).
	In physics, the FKPP equation plays a central role in describing \textit{reaction-diffusion systems}, where a \textit{reaction} process interacts with a \textit{diffusion} mechanism.
	The diffusion term \( \frac{\partial^2 u}{\partial x^2} \) accounts for the spatial spreading of the population or substance while the nonlinear term $f$ accounts for the growth dynamics (reaction).
	FKPP equation is particularly well-known for its travelling wave solutions, which predict the propagation of wavefronts with a constant velocity. 
	One of the key features of the FKPP equation is its ability to predict the speed, shape and stability of these travelling wavefronts, which are determined by the interplay between diffusion and reaction. The ability to predict these properties, together with the equation's versatility in modelling diverse processes from very different field, has established FKPP equations as a cornerstone in the study of nonlinear wave propagation.

To derive the FKPP equation we start from the Quantum Boltzmann equation for each iteration $n$ and consider the following parametrization for the electronic distribution function $F_n$:
\begin{align}
	\label{parametrization_F}
	F_n (\omega,t) \equiv f_{\beta_f}(\omega) +u_n(t) [f_{\beta_i}(\omega)-f_{\beta_f}(\omega)]
\end{align}
Here, $f_{\beta}(\omega)$ is the Fermi distribution at inverse temperature $\beta$ ($\beta_i$ and $\beta_f$ are the initial and final inverse temperatures of the system, respectively) and $u_n(t)$ is a variable whose equation of motion will be determined in this Section.
By taking the derivative in $\omega=0$ of Eq.~\eqref{parametrization_F}, and by sustitutiong the expression $\partial_\omega f_\beta(\omega)|_{\omega=0}=-\frac{1}{4}\beta$ into Eq.~\eqref{parametrization_F} , we get:
\begin{align}
	\label{eqn:beta_n}
	\beta_n(t)=\beta_f +u_n(t)(\beta_i-\beta_f)
\end{align}
This means that $u_n(t)$ is a parametrization for $\beta_n(t)$:
If $u_n=1$, then $\beta_n=\beta_i$; if $u_n=0$, then $\beta_n=\beta_f$.

The scattering integral $I$ in Eq.~\eqref{QB_eq} is made out of three contributions:
\begin{align}
	\label{full_I}
	I_n(\omega,t)=I_{\Sigma_{U,n}}(\omega,t)+ I_{\Delta_n}(\omega,t)+ I_{\Gamma,n}(\omega,t)
\end{align}
Let us start from the first contribution, $I_{\Sigma_U}$, given by the local electron-electron scattering. 
Since $\Sigma^R-\Sigma^A = \Sigma^>-\Sigma^<$, $I_{\Sigma_U}$ in Eq.~\eqref{scatt_int} can be rewritten as
\begin{align}
	\label{scatt_int_les_n}
	I_{\Sigma_{U,n}}(\omega,t) = (-i) \{ \Sigma_n^<(\omega,t)[1-F_n(\omega,t)] + \Sigma_n^>(\omega,t) F_n(\omega,t)\}
\end{align}
In SPT, the the lesser component of the self-energy $\Sigma_{U}$ reads:
\begin{align}
	\label{Sigma_les}
	\Sigma_U^<(\omega) &= U^2(t) \mathcal{F}\{G^<(t)[-G^>(t)^\dag]G^<(t)\} \nonumber \\
	&=U^2(t) \mathcal{F}\{G^<(t)\} * \mathcal{F}\{[-G^>(t)^\dag]\} * \mathcal{F}\{G^<(t)\}
\end{align}
where $G$ is the full (interacting) Green's function.
From the definition of Fourier transform, $\mathcal{F}\{G(t)\}= \frac{1}{2 \pi}\int \mathrm{d} \omega e^{-i \omega t} F(\omega)$, it follows that 
$\mathcal{F}\{- [G^<(t)]^\dag\} = - [G^<(-\omega)]^\dag$. 
In the non-perturbative QBE, when solving the impurity problem, the system is thought to be in a TTI state, thus we can apply the TTI version of the fluctuation-dissipation theorem and write:
\begin{align}
	\label{G_les_gre_n}
	G_n^<(\omega,t)&= i 2 \pi A_n(\omega,t) F_n(\omega,t) \nonumber \\
	G_n^>(\omega,t)&= -i 2 \pi A_n(\omega,t) [1-F_n(\omega,t)] 
\end{align}
According to Eq.~\eqref{G_les_gre_n}, the lesser and greater components of $G$ are fully imaginary, thus $-[G^{<,>}(-\omega)]^*=G^{<,>}(-\omega)$.
It follows that Eq.~\eqref{Sigma_les} (and the corresponding equation for the greater component) becomes:
\begin{align}
	\label{Sigma_les_gre}
	\Sigma_U^{<(>)}(\omega,t) =U^2(t) G^{<(>)}(\omega,t) * G^{>(<)}(-\omega,t) * G^{<(>)}(\omega,t)
\end{align}
where the symbol $*$ indicates convolution.
Let us rewrite explicitly the lesser component of the self-energy:
\begin{align}
	\label{Sigma_les_gre_n}
	\Sigma_{U,n}^{<}(\omega,t) &=U^2(t) (i 2 \pi)^3(-1) \{A_n(\omega,t) F_n(\omega,t) * A_n(-\omega,t) [1-F_n(-\omega,t)] * A_n(\omega,t) F_n(\omega,t) \}
\end{align}
By substituting the expression $f_\beta(\omega)\equiv\frac{1}{1+e^{\beta \omega}}=1-f_\beta(-\omega)$ in Eq.~\eqref{parametrization_F}, we can state that
\begin{align}
	\label{equivalence}
	F_n(\omega,t)=1-F_n(-\omega,t)
\end{align}
We sustitute Eq.~\eqref{equivalence} in Eq.~\eqref{Sigma_les_gre_n} and, since $A_n(\omega)=A_n(-\omega)$ due to the particle-hole invariance of the system, we get:
\begin{align}
	\label{Sigma_les_gre_n1}
	\Sigma_{U,n}^{<}(\omega,t) &=U^2(t) (i 2 \pi)^3(-1) \{A_n(\omega,t) F_n(\omega,t) * A_n(\omega,t) F_n(\omega,t) * A_n(\omega,t) F_n(\omega,t) \}
\end{align}
In a similar manner, for the greater component of the self-energy:
\begin{align}
	\label{Sigma_gre_n}
	\Sigma_{U,n}^{>}(\omega,t) &=U^2(t) (i 2 \pi)^3(-1)^2 \{A_n(\omega,t) [1-F_n(\omega,t)] * A_n(-\omega,t) F_n(-\omega,t) * A_n(\omega,t) [1-F_n(\omega,t)] \}  \\
	& = U^2(t) (i 2 \pi)^3(-1)^2 \{A_n(\omega,t) [1-F_n(\omega,t)] * A_n(\omega,t) [1-F_n(\omega,t)] * A_n(\omega,t) [1-F_n(\omega,t)] \}
	\label{Sigma_gre_n1}
\end{align}
By substituting Eq.~\eqref{Sigma_les_gre_n1} and \eqref{Sigma_gre_n1} into Eq.~\eqref{scatt_int_les_n}, we get:
\begin{align}
	\label{scatt_int_U}
	I_{\Sigma_{U,n}}(\omega,t) &= (-i)U^2(t) (i 2 \pi)^3(-1) [1-F_n(\omega,t)] \{   A_n(\omega,t) F_n(\omega,t) * A_n(\omega,t) F_n(\omega,t) * A_n(\omega,t) F_n(\omega,t) \}+ \nonumber \\
	&(-i)U^2(t) (i 2 \pi)^3(-1)^2 F_n(\omega,t) \{ A_n(\omega,t) [1-F_n(\omega,t)] * A_n(\omega,t) [1-F_n(\omega,t)] * A_n(\omega,t) [1-F_n(\omega,t)]  \}
\end{align}

Now, let us consider the second contribution to the scattering integral, given by the hybridization function $\Delta$:
\begin{align}
	I_{\Delta,n} (\omega,t) = (-i) \{ \Delta_n^<(\omega,t)[1-F_n(\omega,t)] + \Delta_n^>(\omega,t) F_n(\omega,t)\}
\end{align}
Having imposed that $\Delta_n(\omega,t)=t_h^2 G_{n-1}(\omega,t)$ in Eq.~\eqref{Deltan1}, we get:
\begin{align}
	\label{I_Deltann}
	I_{\Delta,n} (\omega,t) = (-i)t_h^2 \{ G_{n-1}^<(\omega,t)[1-F_n(\omega,t)] + G_{n-1}^>(\omega,t) F_n(\omega,t)\}
\end{align}
By substituting Eqs.\eqref{G_les_gre_n} int Eq.~\eqref{I_Deltann}, we get:
\begin{align}
	I_{\Delta,n} (\omega,t) &= 2 \pi t_h^2  A_{n-1}(\omega,t)\{ F_{n-1}(\omega,t) -F_{n}(\omega,t)  \} \\
	& = 2 \pi t_h^2 A_{n-1}(\omega,t) [f_{\beta_i}(\omega) - f_{\beta_f}(\omega)] [u_{n-1}(t)-u_n(t)]
	\label{I_Delta_n}
\end{align}
In the end, we consider the third contribution to the scattering integral due to the coupling with the fermionic bath at negative temperature that mimics photoexcitation:
\begin{align}
	I_{\Gamma,n} (\omega,t) = (-i) \{ \Gamma_n^<(\omega,t)[1-F_n(\omega,t)] + \Gamma_n^>(\omega,t) F_n(\omega,t)\}
\end{align}
By taking $\Gamma^{<,>}$ from Eqs.~\eqref{Gamma}, $I_{\Gamma_n}$ can be rewritten as:
\begin{align}
	\label{I_Gamma}
	I_{\Gamma,n}(\omega,t)  
	&= 2 \pi V^2(t)\{ \mathcal{A}_b(\omega-\omega_0) [1-F_n(\omega,t)] - \mathcal{A}_b(\omega+\omega_0) F_n(\omega,t) \} \\
	&= 2 \pi V^2(t)\{ \mathcal{A}_b(\omega-\omega_0) - f_{\beta_f}(\omega)\mathcal{A}_{\text{bath}}(\omega) - u_n(t) [f_{\beta_i}(\omega) - f_{\beta_f}(\omega)] \mathcal{A}_{\text{bath}}(\omega) \}
	\label{I_Gamma_n}
\end{align}
Going back to the full scattering integral in Eq.~\eqref{full_I}, the QBE reads:
\begin{align}
	\label{Boltzmann}
	\partial_t F_n(\omega,t)= I_{\Sigma_{U,n}}(\omega,t)+ I_{\Delta_n}(\omega,t)+ I_{\Gamma,n}(\omega,t) \equiv I_n(\omega,t)
\end{align}
where: $I_{\Sigma_{U,n}}$ comes from Eq.~\eqref{scatt_int_U}; $I_{\Delta_n}$ from Eq.~\eqref{I_Delta_n}, and $I_{\Gamma_n}$  from Eq.~\eqref{I_Gamma_n}.
On the LHS of Eq.~\eqref{Boltzmann}, we substitute the definition of $F_n$ given by Eq.~\eqref{parametrization_F}, and we get:
\begin{align}
	\label{u_n_start}
	\Delta f (\omega) \frac{{d}}{{dt}}  u_n(t)=I_n(\omega,t)
\end{align}
where we have defined $\Delta f (\omega)$ as $ \Delta f (\omega)\equiv f_{\beta_i}(\omega)-f_{\beta_f}(\omega)$.
Now, we integrate the two members of Eq.~\eqref{u_n_start} in frequency from $0$ to $+\infty$.
By doing this:
\begin{align}
	\label{dt_u_n}
	\frac{{d}}{{dt}} u_n(t) \int_0^{+\infty}\Delta f (\omega)  \text{d}\omega =\int_0^{+\infty} I_n(\omega,t)\text{d}\omega 
\end{align}
By defining $K\equiv\int_0^{+\infty}\Delta f (\omega)\text{d}\omega$, the equation can be rewritten as:
\begin{align}
	\label{u_n}
	\frac{{d}}{{dt}} u_n(t) =\frac{1}{K}\int_0^{+\infty} I_n(\omega,t)\text{d}\omega 
\end{align}
Written explicitly:
\begin{align}
	u_n(t+\Delta t) = u_n(t)+ \frac{\Delta t}{K}\int_0^{+\infty} I_n(\omega,t)\text{d}\omega
	\label{eqn:un} 
\end{align}
We recall that, according to Eq.~\eqref{eqn:beta_n}, $\beta_n(t)=\beta_f +u_n(t)(\beta_i-\beta_f)$, thus:
\begin{align}
	\label{beta_u_n}
	\beta_n(t + \Delta t) &= \beta_f +u_n(t+ \Delta t)(\beta_i-\beta_f)  \nonumber \\
	&=\beta_f + [u_n(t)+ \frac{\Delta t}{K}\int_0^{+\infty} I_n(\omega,t)\text{d}\omega] (\beta_i-\beta_f) \nonumber \\
	&= \beta_f + \Delta \beta [u_n(t)+ \frac{\Delta t}{K}\int_0^{+\infty} I_n(\omega,t)\text{d}\omega ] 
\end{align}
where we have defined $\Delta \beta \equiv(\beta_i-\beta_f)$.
When we solve equation \eqref{dt_u_n}, we make two approximations:
\begin{enumerate}
	\item The spectral functions $A_n(\omega,t)$ which appear in $I_{\Sigma_{U,n}}$ in Eq.~\eqref{scatt_int_U} do not depend on the iteration $n$ and on time $t$: $A_n(\omega,t) \equiv A_{\text{QBE}}(\omega,t_i)$.
	\item The $A_{n-1}(\omega,t)$ in $I_{\Delta,n}$ in Eq.~\eqref{I_Delta_n} do not depend on the iteration $n$ and on time $t$ as well: $A_{n-1}(\omega,t) \equiv A_{\text{QBE}}(\omega,t_i)$
\end{enumerate}
where $\text{QBE}$ indicates the spectral function in the full Noneq-DMFT approach and $t_i$ the last timestep of the simulation (if we had chosen another time, e.g. $t_f$, the initial time, the results wouldn't have changed considerably).
These approximations do not affect considerably the $\beta_n$ which comes form Eq.~\eqref{beta_u_n}, but allow us to write an explicit expression of $I_n$ in terms of powers of $u_n$.

With these approximations, the scattering integral $I_n(\omega,t)$ in Eq.~\eqref{full_I} can be written explicitly as a functional of $u_n$: $I_n(\omega,t) \to I[u_n(t)]$. Indeed:
\begin{itemize}
	\item $I_{\Gamma,n}$ in Eq.~\eqref{I_Gamma_n} is already in the form $I[u_n(t)]$:
	\begin{align}
		I_{\Gamma,n}(\omega,t) = a(t)+b(t) u_n(t)
	\end{align}
	with $a(t)\equiv 2 \pi V^2(t) [\mathcal{A}_b(\omega-\omega_0) - f_{\beta_f}(\omega)\mathcal{A}_{\text{bath}}(\omega)]$ and $b(t) \equiv -2 \pi V^2(t) [f_{\beta_i}(\omega) - f_{\beta_f}(\omega)] \mathcal{A}_{\text{bath}}(\omega)$.
	\item The same for $I_{\Delta,n}$ in Eq.~\eqref{I_Delta_n} :
	\begin{align}
		I_{\Delta,n} (\omega,t) =c [u_{n-1}(t)-u_n(t)]
	\end{align}
	where $c \equiv 2 \pi t_h^2 A_{n-1}(\omega,t) [f_{\beta_i}(\omega) - f_{\beta_f}(\omega)]$ that after the approximation $A_{n-1}(\omega,t) \equiv A_{\text{QBE}}(\omega,t_i)$ becomes $c \equiv 2 \pi t_h^2 A_{\text{QBE}}(\omega,t_i) [f_{\beta_i}(\omega) - f_{\beta_f}(\omega)]$.
	\item  For what concerns $I_{\Sigma,n}$ in Eq.~\eqref{scatt_int_U}, we recall that the function $F_n(\omega,t)$ is parametrized, according to Eq.~\eqref{parametrization_F}, as $F_n (\omega,t) \equiv f_{\beta_f}(\omega) +u_n(t)\Delta f(\omega)$.
	By substituting $F_n (\omega,t) \equiv f_{\beta_f}(\omega) +u_n(t)\Delta f(\omega)$ in Eq.~\eqref{scatt_int_U}, and by making the approximation $A_n(\omega,t) \equiv A_{\text{QBE}}(\omega,t_i) \equiv A$, we see that $I_{\Sigma,n}$ reads:
	\begin{align}
		I_{\Sigma,n}=d + e u_n(t)+f u_n^2(t)+g u_n^3(t) + h u_n^4(t) 
	\end{align}
	Let us write the coefficients $d,\ldots,h$ explicitly:
	\begin{align}
		d(\omega) & = ( A f_{\beta_f} *A f_{\beta_f} * A f_{\beta_f}  )(\omega) - f_{\beta_f}(\omega) ( A*A*A)(\omega)  + \nonumber \\
		&3 f_{\beta_f}(\omega) ( A*A*A f_{\beta_f}  )(\omega) - 3 f_{\beta_f}(\omega) ( A *A f_{\beta_f} * A f_{\beta_f}  )(\omega) =0  \\
		e(\omega) &= 3 ( A \Delta f *A f_{\beta_f} * A f_{\beta_f}  )(\omega) - \Delta f(\omega) ( A *A * A )(\omega) + 3 \Delta f(\omega) ( A *A * A f_{\beta_f} )(\omega)  + \nonumber \\
		&3 f_{\beta_f}(\omega) ( A *A * A \Delta f )(\omega) - 6 f_{\beta_f}(\omega) ( A * A \Delta f * A f_{\beta_f} )(\omega) - 3 \Delta f(\omega) ( A * A f_{\beta_f} * A f_{\beta_f} )(\omega)  \\
		f(\omega) &= 3( A f_{\beta_f}  *A \Delta f * A \Delta f  )(\omega) - 3 f_{\beta_f}(\omega) ( A * A \Delta f * A \Delta f  )(\omega) + \nonumber \\
		& 3 \Delta f (\omega) (A * A * A \Delta f)(\omega) -6 \Delta f (\omega) (A * A f_{\beta_f}  * A \Delta f  )(\omega) \\
		g(\omega) &= (A \Delta f  * A \Delta f  * A \Delta f  )(\omega) - 3 \Delta f (\omega)(A   * A \Delta f  * A \Delta f  )(\omega) \\
		h(\omega) &= ( \Delta f A \Delta f * A \Delta f  * A \Delta f  )(\omega) - ( \Delta f A \Delta f * A \Delta f  * A \Delta f  )(\omega) \equiv 0 
	\end{align}
	The zero-th order coefficient in $u_n$, $d$, and the fourth-order coefficient in $u_n$, $h$, turn out to be zero. $d$ is null because it involves only equilibrium distribution functions $f_\beta$, while $h$ is identically zero because it is the difference between two terms which are equal.
	The non-zero coefficients $e,f,g$, which depend only on $\omega$, 
	are linked by the relation
	\begin{align}
		e+f+g=0    
	\end{align}
	In fact, $u_n(t)=1$ corresponds to $F_n(\omega,t)=f_{\beta_i}(\omega)$; since the system is there at the initial equilibrium temperature, the scattering integral $I_{\Sigma,n}=0$.
\end{itemize}
By putting everything together, the scattering integral $I_n(\omega,t)$ in Eq.~\eqref{full_I} can be written as:
\begin{align}
	I_n [u_n(t)] &= a+b u_n(t)+ c(u_{n-1}(t)-u_n(t))+e u_n(t)+f u_n^2(t)+g u_n^3(t) 
	\label{FKPP}
\end{align}

\begin{figure*}
	\centering \includegraphics[width=\mygraphicwidth]{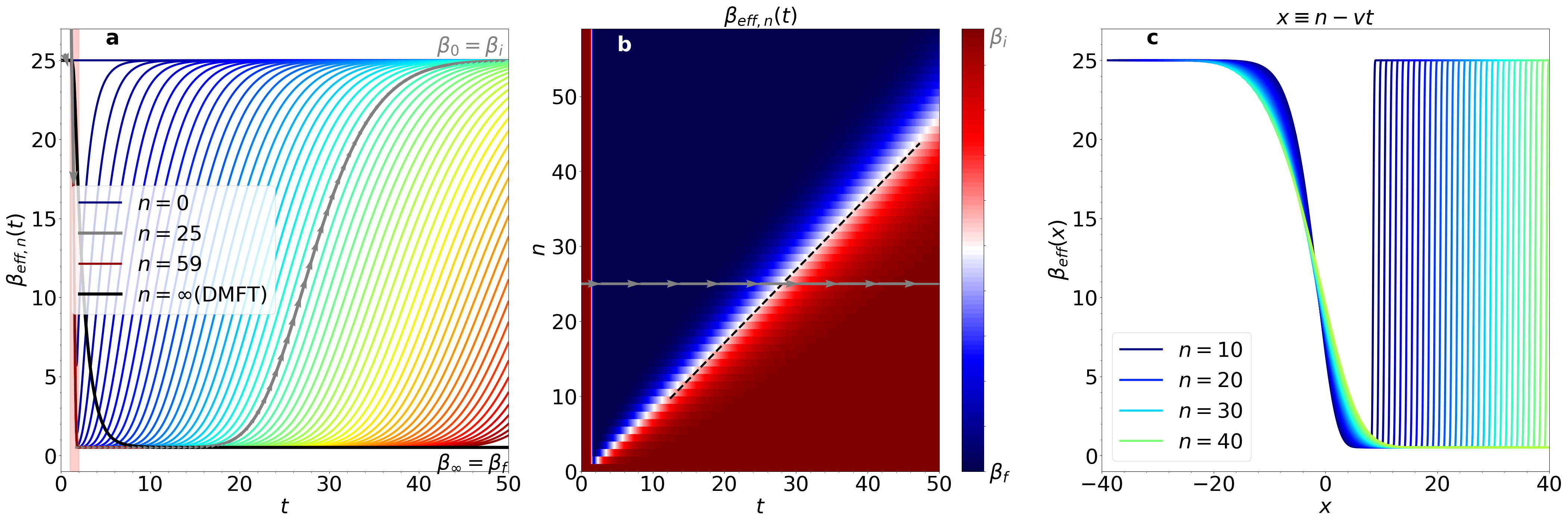}
	\caption{ {\bf Thermalization Front for the Effective Temperature in FKPP}: 
		Panel \textbf{a-c} describe the dynamics of the effective temperature for each iteration $n$, compared to the fully self-consistent DMFT solution (black line) as obtained  from the FKPP equation~(\ref{eqn:FKPP}).
		In Panel \textbf{b}, dashed black line, we plot the velocity of the front in FKPP. 
		In Panel \textbf{c}, we rescale the data for the effective temperature and demonstrate a travelling wave solution, with front velocity $v$.
	}
	\label{fig:DMFT_front_2}
\end{figure*}

By substituting Eq.~\eqref{FKPP} in Eq.~\eqref{u_n} for $u_n$, we get:
\begin{align}
	\label{eqn:FKPP1}
	\dot{u}_n(t) &= C [u_{n-1}(t)-u_n(t) ]+ E u_n(t)+F u_n^2(t)+G u_n^3(t) + A(t) + B(t) u_n (t)
	\nonumber \\
	&=
	C [u_{n-1}(t)-u_n(t) ] + \Phi [u_n(t)] +\eta_n(t)
\end{align}
where 
\begin{align}
	\label{eqn:Phi}
	\Phi (u_n) &\equiv  E u_n+F u_n^2+G u_n^3 \\
	\eta_n(t) &\equiv A(t) + B(t) u_n (t) 
\end{align}
and the capital coefficients $A,\ldots,G$ are related to the lowercase ones by the relation:
\begin{align}
	\label{eqn:Acoeff}
	A \equiv \frac{1}{K} \int_0^{+\infty} a(\omega, \cdot) \text{d}\omega
\end{align}
We can recognize an FKPP structure in Eq.~\eqref{eqn:FKPP1} where there it appears the term proportional to $u_n-u_{n-1}$ through the coefficient $C$ and the terms nonlinear in $u_n$, $\Phi (u_n)$. Solving it for $u_n(t)$, we obtain a sharp front propagating ballistically (Fig.~\ref{fig:DMFT_front_2} (a)). Moreover, the velocity of the front (Panel \textbf{b}) is captured by Eq.~(\ref{eqn:FKPP}) of the main text, as well as the shape of the front (Panel \textbf{c}).


\subsection{Linearization of FKPP Dynamics and Critical Behavior}
\label{sec:FKPP}
In this section we discuss the properties of the non-linear ordinary differential equation (ODE) Eq.~\eqref{eqn:FKPP1} for $u_n$, and we look for travelling-wave solutions of the type~\cite{brunetderrida2015}:
\begin{align}
	\label{eqn:travelling_n}
	u_n(t)=u_{n+1} \big ( t+\frac{1}{v} \big )  
\end{align}
where $v$ is the wave velocity.
By substituting Eq.~\eqref{eqn:travelling_n} into Eq.~\eqref{eqn:FKPP1}, where we set $\eta_n=0$ since it represents just the initial trigger for the time evolution, we get:
\begin{align}
	\label{eqn:FKPP2}
	\dot{u}_n(t) &= C \big [u_{n} \big ( t+\frac{1}{v}\big )-u_n(t) \big ]+ \Phi[u_n(t)]
\end{align}
We define a continuous family $W_v$ of travelling wave solutions:
\begin{align}
	u_n(t) = W_v (n-vt)
\end{align}
where the variable $x$ is $x \equiv n - vt$.
Since
\begin{align}
	\frac{{d}}{{dt}} u_n(t) &= \partial_t W_v (n-vt) = \frac{{d}}{{dx}} W_v (x) \partial_t x =  W_v'(x) (-v)  \nonumber \\
	u_{n} \big ( t+\frac{1}{v}\big ) &= W_v \big [ n- v\big( t +\frac{1}{v} \big ) \big] = W_v (n-vt-1) =W_v (x-1)
\end{align}
Eq.~\eqref{eqn:FKPP2} satisfies the ODE
\begin{align}
	\label{eqn:FKPP3}
	(-v) W_v'(x) = C [W_v (x-1) - W_v(x) ] + \Phi [W_v(x)],
\end{align}
where $W_v'(x) \equiv \frac{{d}}{{dt}} W_v(x)$, with the boundary conditions $W_v(-\infty) =1 $ and $W_v(+\infty) =0$. 
By linearizing Eq.~\eqref{eqn:FKPP3} for small $W_v$ (when $x$ is large)
\begin{align}
	\label{eqn:FKPP4}
	(-v) W_v'(x) = C [W_v (x-1) - W_v(x) ] + E W_v(x),
\end{align}
one can see that $W_v(x)$ vanishes exponentially as $x\to \infty$:
\begin{align}
	\label{eqn:gamma}
	W_v(x) \sim e^{-\gamma x}
\end{align}
The coefficient $E$ in Eq.~\eqref{eqn:FKPP4} comes from the liner expansion of Eq.~\eqref{eqn:Phi}.
By substituting the expression \eqref{eqn:gamma} into the ODE \eqref{eqn:FKPP4}, we get the dispersion relation $v=v(\gamma)$:
\begin{align}
	\label{eqn:dispersion_relation}
	v(\gamma) =\frac{E-C+Ce^\gamma}{\gamma}
\end{align}
The function $v(\gamma)$  has the property $ v(\gamma) \to \infty$ both for $ \gamma \to 0^+$ and $\gamma \to + \infty$. In between, it has a single minimum in correspondence of the critical value of $\gamma$,  $\gamma_c$:
\begin{align}
	v_c \equiv \min_\gamma v(\gamma)  =  v(\gamma_c) 
\end{align}

Eq. ~\eqref{eqn:FKPP4} admits travelling wave solutions for all velocities $v$ larger or equal than the critical velocity $v_c$~\cite{aron2023kinetics}. This means that for all $v \geq v_c$ there exists functions $W_v$ such that $W_v(n-vt)$ is a solution of Eq.~\eqref{eqn:FKPP4}.
Following Ref.~\cite{brunetderrida2015}, we can state it more explicitly:
\begin{itemize}
	\item For $0< v <v_c$, the solutions of $v(\gamma)=v$ are complex, so $W_v$ oscillates around $x=0$ and goes to zero as $x \to \infty$.
	\item For $v>v_c$, the travelling wave $W_v$ is monotonically decreasing and decays for large $x$ as:
	\begin{align}
		W_v (x) \approx A  e^{-\gamma_1 x}
	\end{align}
	with $A>0$. $\gamma_1$ is the smallest of the two solutions $\gamma_{1,2}$ of the equation $v(\gamma) = v$, with $v$ constant $v \in (0,v_c)$.
	\item For $v=v_c$, the two roots of $v(\gamma)=v_c$ coincide at $\gamma_c$. The corresponding travelling wave is monotonically decreasing and decays for large $x$ as 
	\begin{align}
		\label{eqn:Wc}
		W_{v_c}(x) \approx A x e^{-\gamma_c x}
	\end{align}
	with $A>0$. By substituting Eq.~\eqref{eqn:Wc} into the ODE~\eqref{eqn:FKPP4}, one finds that 
	\begin{align}
		v_c = C e^{\gamma_c}
	\end{align}
	where $\gamma_c$ is the $\gamma$ which corresponds to the minimum of $v(\gamma)$ in Eq.~\eqref{eqn:dispersion_relation}.
\end{itemize}


\begin{figure*}
	\centering \includegraphics[width=0.8\textwidth]{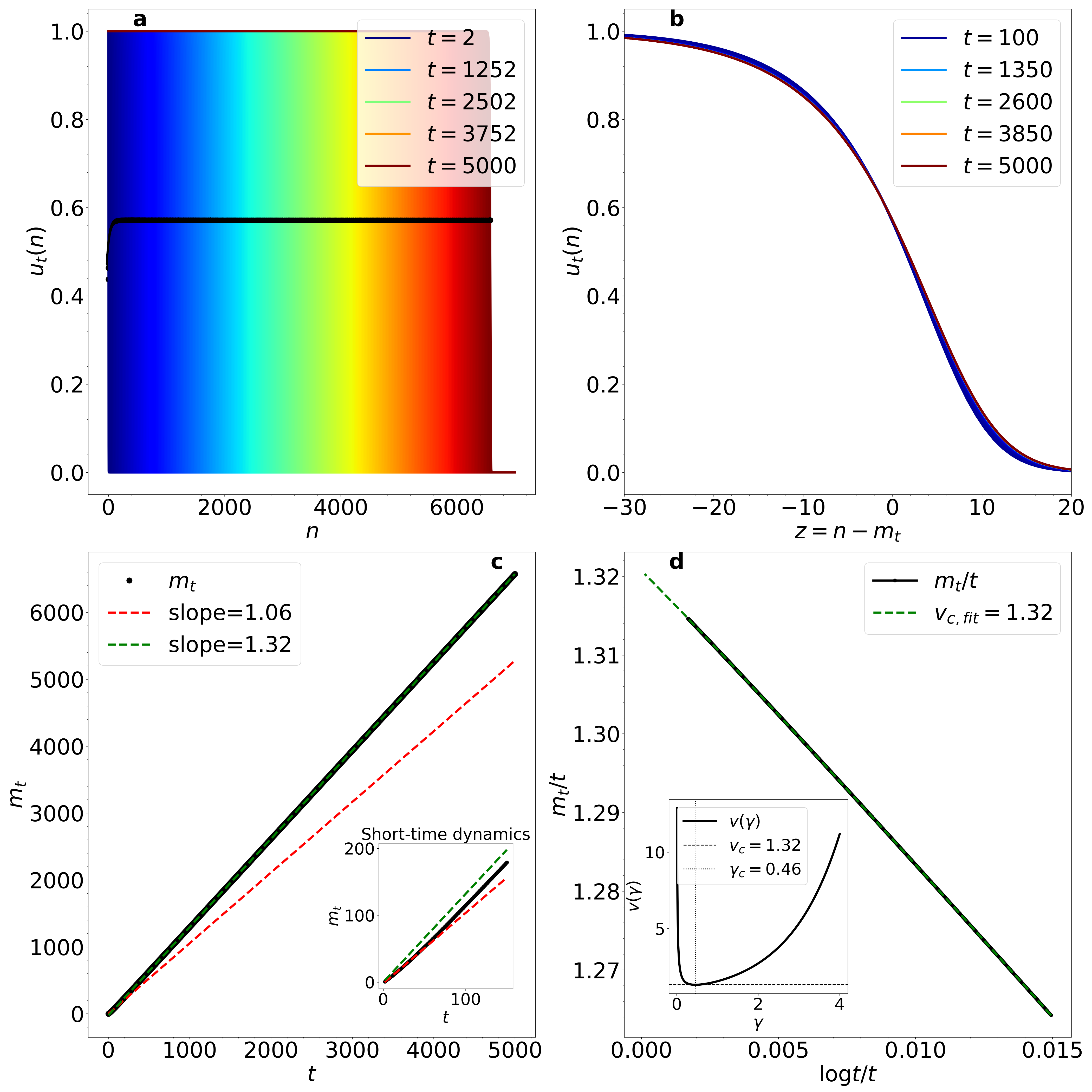}
	\caption{ {\bf Critical behaviour in the FKPP propagation}: 
		Panel \textbf{a}: Function $u_t(n)$ versus $n$. For each coloured curve, which corresponds to a given time $t$, the value $m_t$ is extracted as $m_t = \int_{0}^{+\infty} u_t(n) \text{d} n $. This value corresponds to the position of the wave-front at a given time $t$ (dotted black line).  Panel \textbf{b}: $u_t(n)$ versus $n-m_t$ illustrates that all the rescaled functions, for $t\geq 100$, lie on top of each other.  Panel \textbf{c}: $m_t$ versus $t$ shows that the system, for short times, has a velocity which is  $v_1 \approx 1.06$ (see inset), while for longer times it approaches the velocity $v_2 \approx 1.32$. The latter is exactly the critical velocity one expects for the front as $t \to \infty$ from the FKPP analysis. In fact, from the dispersion relation $v(\gamma)$ (inset panel \textbf{d}), we find that the minimum of the curve is for $v_c\approx1.32$ and $\gamma_c \approx 0.46$. 
		Panel \textbf{d}: Fit of $m_t/t$ versus $ \frac{\log{t}}{t}$. From the fit, $v_{c,\text{fit}} \approx 1.32 =v_c$.
	}
	\label{fig:DMFT_front_3}
\end{figure*}
In order to test the validity of the FKPP-relations extracted above, we make an analysis on the position and the shape of the wave-front.
In Fig.~\ref{fig:DMFT_front_3} \textbf{a}, we display the curve $u_t(n)$, as extracted from Eq.~\eqref{eqn:un}, versus the iteration $n$. There, each coloured curve represents a given time step $t$. For each of them, the position of the front, $m_t$, is calculated as $m_t = \int_{0}^{+\infty} u_t(n) \text{d} n$ (dotted black line). By rescaling the curves with respect to $n-m_t$, one can appreciate their collapse in a stable front for all $t \geq 100$, reminiscent of a travelling wave equation (panel \textbf{b}).
In panel \textbf{c}, we plot the position of the front, taken from panel \textbf{a}, versus time. 
$m_t$ versus $t$ shows that the system, for short times, has a velocity which is  $v_1 \approx 1.06$ (see inset), while for longer times it approaches the velocity $v_2 \approx 1.32$. 
The latter is exactly the critical velocity one expects for the front as $t \to \infty$ from the FKPP analysis. In fact, from the dispersion relation $v(\gamma)$ in Eq.~\eqref{eqn:dispersion_relation}  (inset panel \textbf{d}), the minimum of the curve $v(\gamma)$ is found at $\gamma_c \approx 0.46$, which corresponds to $v_c(\gamma_c)\approx1.32$.
From the FKPP analysis~\cite{aron2023kinetics},  we indeed expect $m_t \sim v_c t$ as $t \to \infty$. More precisely, the asymptotic position of the front is
\begin{align}
	m_t = v_c t - B \log{t} +a -\frac{E}{\sqrt{t}} + \mathcal{O}\big ( \frac{\log{t}}{t} \big ) \,\,\,\, \text{as} \,\,\,\, t \to \infty
	\label{eqn:m_t}
\end{align}
By fitting $m_t$ from panel \textbf{a}  with the expression~\eqref{eqn:m_t}, we find $v_{c,\text{fit}} \approx 1.32$. This value coincides with the value of the critical velocity that one expects from Eq.~\eqref{eqn:dispersion_relation}, thus confirming that the long-time behaviour of the system follows the FKPP-equations.
In panel \textbf{d}, we show $m_t/t$, as taken from panel \textbf{a}, versus $\log{t}/t$ (dotted black line). The fit of $m_t$ with the expression in Eq.~\eqref{eqn:m_t} (dashed green line), with $v_c \equiv v_{c,\text{fit}}$, perfectly reproduces the data.

\subsection{Comparison with DMFT}

In order to compare the DMFT results obtained with the QB equation (see Fig.~\ref{fig:DMFT_front} main text) with the predictions from the FKPP analysis discussed in the previous section, we proceed as following. We first extract the velocity from the front in the short-time dynamics of the DMFT data shown in the main text, Fig.~\ref{fig:DMFT_front}(b). This velocity, $v=1.06$, turns out to be equal to the prediction of the FKPP equation at short times, see inset of Fig~\ref{fig:DMFT_front_3}(c). As we know from the previous analysis, this velocity drifts at long times towards a larger value $v_c=1.32$, which coincides precisely with the critical velocity obtained from the minimum of the dispersion relation in Eq.~\eqref{eqn:dispersion_relation}.
For an FKPP front, the behaviour of $v(\gamma)$ around the minimum at $\gamma=\gamma_c$ controls many properties of the propagating front, including the position and the shape of the front~\cite{aron2023kinetics}. We therefore fix $\gamma_c$ by taking the minimum of the dispersion relation $v(\gamma)$ in Eq.~\eqref{eqn:dispersion_relation},
according to $ v_c \equiv \min_\gamma v(\gamma)  =  v(\gamma_c) $ (see inset of Fig~\ref{fig:DMFT_front_3}(d)).
We then substitute $\gamma_c$ into Eq.~\eqref{eqn:Wc} in order to get the long-$x$ behaviour of $W_{v_c}(x)$. Finally, we compare the behaviour of $W_{v_c}(x)$ with the rescaled SPT curves of $\beta_n(x)$ in Fig.~\ref{fig:DMFT_front} \textbf{c}. We see that the long-$x$ behaviour of $\beta_n(x)$ is well fitted by the $W_{v_c}(x)$  which comes form Eq.~\eqref{eqn:Wc}(red dotted curve). We point out that, before comparing the two curves, we rescaled the $W_{v_c}(x)$ data in Eq.~\eqref{eqn:Wc} according to relation~\eqref{eqn:beta_n}:
\begin{align}
	\label{eqn:beta_FKPP}
	\beta_{\text{FKPP}}(x) = \beta_f + W_{v_c}(x)(\beta_i-\beta_f)
\end{align}
The good fitting of the data in Fig.~\ref{fig:DMFT_front} \textbf{c} with the function $W_{v_c}(x) \approx A x e^{-\gamma_c x}$ confirms the FKPP nature of the dynamics of the travelling waves propagating in our system.


\begin{thebibliography}{45}%
	\makeatletter
	\providecommand \@ifxundefined [1]{%
		\@ifx{#1\undefined}
	}%
	\providecommand \@ifnum [1]{%
		\ifnum #1\expandafter \@firstoftwo
		\else \expandafter \@secondoftwo
		\fi
	}%
	\providecommand \@ifx [1]{%
		\ifx #1\expandafter \@firstoftwo
		\else \expandafter \@secondoftwo
		\fi
	}%
	\providecommand \natexlab [1]{#1}%
	\providecommand \enquote  [1]{``#1''}%
	\providecommand \bibnamefont  [1]{#1}%
	\providecommand \bibfnamefont [1]{#1}%
	\providecommand \citenamefont [1]{#1}%
	\providecommand \href@noop [0]{\@secondoftwo}%
	\providecommand \href [0]{\begingroup \@sanitize@url \@href}%
	\providecommand \@href[1]{\@@startlink{#1}\@@href}%
	\providecommand \@@href[1]{\endgroup#1\@@endlink}%
	\providecommand \@sanitize@url [0]{\catcode `\\12\catcode `\$12\catcode
		`\&12\catcode `\#12\catcode `\^12\catcode `\_12\catcode `\%12\relax}%
	\providecommand \@@startlink[1]{}%
	\providecommand \@@endlink[0]{}%
	\providecommand \url  [0]{\begingroup\@sanitize@url \@url }%
	\providecommand \@url [1]{\endgroup\@href {#1}{\urlprefix }}%
	\providecommand \urlprefix  [0]{URL }%
	\providecommand \Eprint [0]{\href }%
	\providecommand \doibase [0]{https://doi.org/}%
	\providecommand \selectlanguage [0]{\@gobble}%
	\providecommand \bibinfo  [0]{\@secondoftwo}%
	\providecommand \bibfield  [0]{\@secondoftwo}%
	\providecommand \translation [1]{[#1]}%
	\providecommand \BibitemOpen [0]{}%
	\providecommand \bibitemStop [0]{}%
	\providecommand \bibitemNoStop [0]{.\EOS\space}%
	\providecommand \EOS [0]{\spacefactor3000\relax}%
	\providecommand \BibitemShut  [1]{\csname bibitem#1\endcsname}%
	\let\auto@bib@innerbib\@empty
	\bibitem [{\citenamefont {Polkovnikov}\ \emph {et~al.}(2011)\citenamefont
		{Polkovnikov}, \citenamefont {Sengupta}, \citenamefont {Silva},\ and\
		\citenamefont {Vengalattore}}]{polkovnikov2011colloquium}%
	\BibitemOpen
	\bibfield  {author} {\bibinfo {author} {\bibfnamefont {A.}~\bibnamefont
			{Polkovnikov}}, \bibinfo {author} {\bibfnamefont {K.}~\bibnamefont
			{Sengupta}}, \bibinfo {author} {\bibfnamefont {A.}~\bibnamefont {Silva}},\
		and\ \bibinfo {author} {\bibfnamefont {M.}~\bibnamefont {Vengalattore}},\
	}\bibfield  {title} {\bibinfo {title} {Colloquium: Nonequilibrium dynamics of
			closed interacting quantum systems},\ }\href
	{https://doi.org/10.1103/RevModPhys.83.863} {\bibfield  {journal} {\bibinfo
			{journal} {Rev. Mod. Phys.}\ }\textbf {\bibinfo {volume} {83}},\ \bibinfo
		{pages} {863} (\bibinfo {year} {2011})}\BibitemShut {NoStop}%
	\bibitem [{\citenamefont {Gogolin}\ and\ \citenamefont
		{Eisert}(2016)}]{Gogolin_2016}%
	\BibitemOpen
	\bibfield  {author} {\bibinfo {author} {\bibfnamefont {C.}~\bibnamefont
			{Gogolin}}\ and\ \bibinfo {author} {\bibfnamefont {J.}~\bibnamefont
			{Eisert}},\ }\bibfield  {title} {\bibinfo {title} {Equilibration,
			thermalisation, and the emergence of statistical mechanics in closed quantum
			systems},\ }\href {https://doi.org/10.1088/0034-4885/79/5/056001} {\bibfield
		{journal} {\bibinfo  {journal} {Reports on Progress in Physics}\ }\textbf
		{\bibinfo {volume} {79}},\ \bibinfo {pages} {056001} (\bibinfo {year}
		{2016})}\BibitemShut {NoStop}%
	\bibitem [{\citenamefont {D'Alessio}\ \emph {et~al.}(2016)\citenamefont
		{D'Alessio}, \citenamefont {Kafri}, \citenamefont {Polkovnikov},\ and\
		\citenamefont {Rigol}}]{dalessio2016from}%
	\BibitemOpen
	\bibfield  {author} {\bibinfo {author} {\bibfnamefont {L.}~\bibnamefont
			{D'Alessio}}, \bibinfo {author} {\bibfnamefont {Y.}~\bibnamefont {Kafri}},
		\bibinfo {author} {\bibfnamefont {A.}~\bibnamefont {Polkovnikov}},\ and\
		\bibinfo {author} {\bibfnamefont {M.}~\bibnamefont {Rigol}},\ }\bibfield
	{title} {\bibinfo {title} {From quantum chaos and eigenstate thermalization
			to statistical mechanics and thermodynamics},\ }\href
	{https://doi.org/10.1080/00018732.2016.1198134} {\bibfield  {journal}
		{\bibinfo  {journal} {Advances in Physics}\ }\textbf {\bibinfo {volume}
			{65}},\ \bibinfo {pages} {239} (\bibinfo {year} {2016})},\ \Eprint
	{https://arxiv.org/abs/https://doi.org/10.1080/00018732.2016.1198134}
	{https://doi.org/10.1080/00018732.2016.1198134} \BibitemShut {NoStop}%
	\bibitem [{\citenamefont {Kaufman}\ \emph {et~al.}(2016)\citenamefont
		{Kaufman}, \citenamefont {Tai}, \citenamefont {Lukin}, \citenamefont
		{Rispoli}, \citenamefont {Schittko}, \citenamefont {Preiss},\ and\
		\citenamefont {Greiner}}]{kaufman2016quantum}%
	\BibitemOpen
	\bibfield  {author} {\bibinfo {author} {\bibfnamefont {A.~M.}\ \bibnamefont
			{Kaufman}}, \bibinfo {author} {\bibfnamefont {M.~E.}\ \bibnamefont {Tai}},
		\bibinfo {author} {\bibfnamefont {A.}~\bibnamefont {Lukin}}, \bibinfo
		{author} {\bibfnamefont {M.}~\bibnamefont {Rispoli}}, \bibinfo {author}
		{\bibfnamefont {R.}~\bibnamefont {Schittko}}, \bibinfo {author}
		{\bibfnamefont {P.~M.}\ \bibnamefont {Preiss}},\ and\ \bibinfo {author}
		{\bibfnamefont {M.}~\bibnamefont {Greiner}},\ }\bibfield  {title} {\bibinfo
		{title} {Quantum thermalization through entanglement in an isolated many-body
			system},\ }\href {https://doi.org/10.1126/science.aaf6725} {\bibfield
		{journal} {\bibinfo  {journal} {Science}\ }\textbf {\bibinfo {volume}
			{353}},\ \bibinfo {pages} {794} (\bibinfo {year} {2016})},\ \Eprint
	{https://arxiv.org/abs/https://www.science.org/doi/pdf/10.1126/science.aaf6725}
	{https://www.science.org/doi/pdf/10.1126/science.aaf6725} \BibitemShut
	{NoStop}%
	\bibitem [{\citenamefont {Kranzl}\ \emph {et~al.}(2023)\citenamefont {Kranzl},
		\citenamefont {Lasek}, \citenamefont {Joshi}, \citenamefont {Kalev},
		\citenamefont {Blatt}, \citenamefont {Roos},\ and\ \citenamefont
		{Yunger~Halpern}}]{kranzl2023experimental}%
	\BibitemOpen
	\bibfield  {author} {\bibinfo {author} {\bibfnamefont {F.}~\bibnamefont
			{Kranzl}}, \bibinfo {author} {\bibfnamefont {A.}~\bibnamefont {Lasek}},
		\bibinfo {author} {\bibfnamefont {M.~K.}\ \bibnamefont {Joshi}}, \bibinfo
		{author} {\bibfnamefont {A.}~\bibnamefont {Kalev}}, \bibinfo {author}
		{\bibfnamefont {R.}~\bibnamefont {Blatt}}, \bibinfo {author} {\bibfnamefont
			{C.~F.}\ \bibnamefont {Roos}},\ and\ \bibinfo {author} {\bibfnamefont
			{N.}~\bibnamefont {Yunger~Halpern}},\ }\bibfield  {title} {\bibinfo {title}
		{Experimental observation of thermalization with noncommuting charges},\
	}\href {https://doi.org/10.1103/PRXQuantum.4.020318} {\bibfield  {journal}
		{\bibinfo  {journal} {PRX Quantum}\ }\textbf {\bibinfo {volume} {4}},\
		\bibinfo {pages} {020318} (\bibinfo {year} {2023})}\BibitemShut {NoStop}%
	\bibitem [{\citenamefont {Berry}(1977)}]{MVBerry_1977}%
	\BibitemOpen
	\bibfield  {author} {\bibinfo {author} {\bibfnamefont {M.~V.}\ \bibnamefont
			{Berry}},\ }\bibfield  {title} {\bibinfo {title} {Regular and irregular
			semiclassical wavefunctions},\ }\href
	{https://doi.org/10.1088/0305-4470/10/12/016} {\bibfield  {journal} {\bibinfo
			{journal} {Journal of Physics A: Mathematical and General}\ }\textbf
		{\bibinfo {volume} {10}},\ \bibinfo {pages} {2083} (\bibinfo {year}
		{1977})}\BibitemShut {NoStop}%
	\bibitem [{\citenamefont {Deutsch}(1991)}]{Deutsch91}%
	\BibitemOpen
	\bibfield  {author} {\bibinfo {author} {\bibfnamefont {J.~M.}\ \bibnamefont
			{Deutsch}},\ }\bibfield  {title} {\bibinfo {title} {Quantum statistical
			mechanics in a closed system},\ }\href
	{https://doi.org/10.1103/PhysRevA.43.2046} {\bibfield  {journal} {\bibinfo
			{journal} {Phys. Rev. A}\ }\textbf {\bibinfo {volume} {43}},\ \bibinfo
		{pages} {2046} (\bibinfo {year} {1991})}\BibitemShut {NoStop}%
	\bibitem [{\citenamefont {Srednicki}(1994)}]{Srednicki_ETH}%
	\BibitemOpen
	\bibfield  {author} {\bibinfo {author} {\bibfnamefont {M.}~\bibnamefont
			{Srednicki}},\ }\bibfield  {title} {\bibinfo {title} {Chaos and quantum
			thermalization},\ }\href {https://doi.org/10.1103/PhysRevE.50.888} {\bibfield
		{journal} {\bibinfo  {journal} {Phys. Rev. E}\ }\textbf {\bibinfo {volume}
			{50}},\ \bibinfo {pages} {888} (\bibinfo {year} {1994})}\BibitemShut
	{NoStop}%
	\bibitem [{\citenamefont {Srednicki}(1999)}]{MarkSrednicki_1999}%
	\BibitemOpen
	\bibfield  {author} {\bibinfo {author} {\bibfnamefont {M.}~\bibnamefont
			{Srednicki}},\ }\bibfield  {title} {\bibinfo {title} {The approach to thermal
			equilibrium in quantized chaotic systems},\ }\href
	{https://doi.org/10.1088/0305-4470/32/7/007} {\bibfield  {journal} {\bibinfo
			{journal} {Journal of Physics A: Mathematical and General}\ }\textbf
		{\bibinfo {volume} {32}},\ \bibinfo {pages} {1163} (\bibinfo {year}
		{1999})}\BibitemShut {NoStop}%
	\bibitem [{\citenamefont {Rigol}\ \emph {et~al.}(2008)\citenamefont {Rigol},
		\citenamefont {Dunjko},\ and\ \citenamefont
		{Olshanii}}]{rigol2008thermalization}%
	\BibitemOpen
	\bibfield  {author} {\bibinfo {author} {\bibfnamefont {M.}~\bibnamefont
			{Rigol}}, \bibinfo {author} {\bibfnamefont {V.}~\bibnamefont {Dunjko}},\ and\
		\bibinfo {author} {\bibfnamefont {M.}~\bibnamefont {Olshanii}},\ }\bibfield
	{title} {\bibinfo {title} {Thermalization and its mechanism for generic
			isolated quantum systems},\ }\href {https://doi.org/10.1038/nature06838}
	{\bibfield  {journal} {\bibinfo  {journal} {Nature}\ }\textbf {\bibinfo
			{volume} {452}},\ \bibinfo {pages} {854} (\bibinfo {year}
		{2008})}\BibitemShut {NoStop}%
	\bibitem [{\citenamefont {Biroli}\ \emph {et~al.}(2010)\citenamefont {Biroli},
		\citenamefont {Kollath},\ and\ \citenamefont
		{L\"auchli}}]{Biroli_Corinna_prl10}%
	\BibitemOpen
	\bibfield  {author} {\bibinfo {author} {\bibfnamefont {G.}~\bibnamefont
			{Biroli}}, \bibinfo {author} {\bibfnamefont {C.}~\bibnamefont {Kollath}},\
		and\ \bibinfo {author} {\bibfnamefont {A.~M.}\ \bibnamefont {L\"auchli}},\
	}\bibfield  {title} {\bibinfo {title} {Effect of rare fluctuations on the
			thermalization of isolated quantum systems},\ }\href
	{https://doi.org/10.1103/PhysRevLett.105.250401} {\bibfield  {journal}
		{\bibinfo  {journal} {Phys. Rev. Lett.}\ }\textbf {\bibinfo {volume} {105}},\
		\bibinfo {pages} {250401} (\bibinfo {year} {2010})}\BibitemShut {NoStop}%
	\bibitem [{\citenamefont {Ikeda}\ \emph {et~al.}(2013)\citenamefont {Ikeda},
		\citenamefont {Watanabe},\ and\ \citenamefont {Ueda}}]{ikeda2013finite}%
	\BibitemOpen
	\bibfield  {author} {\bibinfo {author} {\bibfnamefont {T.~N.}\ \bibnamefont
			{Ikeda}}, \bibinfo {author} {\bibfnamefont {Y.}~\bibnamefont {Watanabe}},\
		and\ \bibinfo {author} {\bibfnamefont {M.}~\bibnamefont {Ueda}},\ }\bibfield
	{title} {\bibinfo {title} {Finite-size scaling analysis of the eigenstate
			thermalization hypothesis in a one-dimensional interacting bose gas},\ }\href
	{https://doi.org/10.1103/PhysRevE.87.012125} {\bibfield  {journal} {\bibinfo
			{journal} {Phys. Rev. E}\ }\textbf {\bibinfo {volume} {87}},\ \bibinfo
		{pages} {012125} (\bibinfo {year} {2013})}\BibitemShut {NoStop}%
	\bibitem [{\citenamefont {Kim}\ \emph {et~al.}(2014)\citenamefont {Kim},
		\citenamefont {Ikeda},\ and\ \citenamefont {Huse}}]{kim2014testing}%
	\BibitemOpen
	\bibfield  {author} {\bibinfo {author} {\bibfnamefont {H.}~\bibnamefont
			{Kim}}, \bibinfo {author} {\bibfnamefont {T.~N.}\ \bibnamefont {Ikeda}},\
		and\ \bibinfo {author} {\bibfnamefont {D.~A.}\ \bibnamefont {Huse}},\
	}\bibfield  {title} {\bibinfo {title} {Testing whether all eigenstates obey
			the eigenstate thermalization hypothesis},\ }\href
	{https://doi.org/10.1103/PhysRevE.90.052105} {\bibfield  {journal} {\bibinfo
			{journal} {Phys. Rev. E}\ }\textbf {\bibinfo {volume} {90}},\ \bibinfo
		{pages} {052105} (\bibinfo {year} {2014})}\BibitemShut {NoStop}%
	\bibitem [{\citenamefont {Reimann}(2016)}]{reiman2016}%
	\BibitemOpen
	\bibfield  {author} {\bibinfo {author} {\bibfnamefont {P.}~\bibnamefont
			{Reimann}},\ }\bibfield  {title} {\bibinfo {title} {Typical fast
			thermalization processes in closed many-body systems},\ }\href
	{https://doi.org/10.1038/ncomms10821} {\bibfield  {journal} {\bibinfo
			{journal} {Nature Communications}\ }\textbf {\bibinfo {volume} {7}},\
		\bibinfo {pages} {10821} (\bibinfo {year} {2016})}\BibitemShut {NoStop}%
	\bibitem [{\citenamefont {Hallam}\ \emph {et~al.}(2019)\citenamefont {Hallam},
		\citenamefont {Morley},\ and\ \citenamefont {Green}}]{hallam2019thelyapunov}%
	\BibitemOpen
	\bibfield  {author} {\bibinfo {author} {\bibfnamefont {A.}~\bibnamefont
			{Hallam}}, \bibinfo {author} {\bibfnamefont {J.~G.}\ \bibnamefont {Morley}},\
		and\ \bibinfo {author} {\bibfnamefont {A.~G.}\ \bibnamefont {Green}},\
	}\bibfield  {title} {\bibinfo {title} {The lyapunov spectra of quantum
			thermalisation},\ }\href {https://doi.org/10.1038/s41467-019-10336-4}
	{\bibfield  {journal} {\bibinfo  {journal} {Nature Communications}\ }\textbf
		{\bibinfo {volume} {10}},\ \bibinfo {pages} {2708} (\bibinfo {year}
		{2019})}\BibitemShut {NoStop}%
	\bibitem [{\citenamefont {Foini}\ and\ \citenamefont
		{Kurchan}(2019)}]{foini2019eigenstate}%
	\BibitemOpen
	\bibfield  {author} {\bibinfo {author} {\bibfnamefont {L.}~\bibnamefont
			{Foini}}\ and\ \bibinfo {author} {\bibfnamefont {J.}~\bibnamefont
			{Kurchan}},\ }\bibfield  {title} {\bibinfo {title} {Eigenstate thermalization
			hypothesis and out of time order correlators},\ }\href
	{https://doi.org/10.1103/PhysRevE.99.042139} {\bibfield  {journal} {\bibinfo
			{journal} {Phys. Rev. E}\ }\textbf {\bibinfo {volume} {99}},\ \bibinfo
		{pages} {042139} (\bibinfo {year} {2019})}\BibitemShut {NoStop}%
	\bibitem [{\citenamefont {Pappalardi}\ \emph {et~al.}(2022)\citenamefont
		{Pappalardi}, \citenamefont {Foini},\ and\ \citenamefont
		{Kurchan}}]{pappalardi2022eigenstate}%
	\BibitemOpen
	\bibfield  {author} {\bibinfo {author} {\bibfnamefont {S.}~\bibnamefont
			{Pappalardi}}, \bibinfo {author} {\bibfnamefont {L.}~\bibnamefont {Foini}},\
		and\ \bibinfo {author} {\bibfnamefont {J.}~\bibnamefont {Kurchan}},\
	}\bibfield  {title} {\bibinfo {title} {Eigenstate thermalization hypothesis
			and free probability},\ }\href
	{https://doi.org/10.1103/PhysRevLett.129.170603} {\bibfield  {journal}
		{\bibinfo  {journal} {Phys. Rev. Lett.}\ }\textbf {\bibinfo {volume} {129}},\
		\bibinfo {pages} {170603} (\bibinfo {year} {2022})}\BibitemShut {NoStop}%
	\bibitem [{\citenamefont {Chan}\ \emph {et~al.}(2022)\citenamefont {Chan},
		\citenamefont {Shivam}, \citenamefont {Huse},\ and\ \citenamefont
		{De~Luca}}]{chan2022manybody}%
	\BibitemOpen
	\bibfield  {author} {\bibinfo {author} {\bibfnamefont {A.}~\bibnamefont
			{Chan}}, \bibinfo {author} {\bibfnamefont {S.}~\bibnamefont {Shivam}},
		\bibinfo {author} {\bibfnamefont {D.~A.}\ \bibnamefont {Huse}},\ and\
		\bibinfo {author} {\bibfnamefont {A.}~\bibnamefont {De~Luca}},\ }\bibfield
	{title} {\bibinfo {title} {Many-body quantum chaos and space-time
			translational invariance},\ }\href
	{https://doi.org/10.1038/s41467-022-34318-1} {\bibfield  {journal} {\bibinfo
			{journal} {Nature Communications}\ }\textbf {\bibinfo {volume} {13}},\
		\bibinfo {pages} {7484} (\bibinfo {year} {2022})}\BibitemShut {NoStop}%
	\bibitem [{\citenamefont {Nandkishore}\ and\ \citenamefont
		{Huse}(2015)}]{Nandkishore2015}%
	\BibitemOpen
	\bibfield  {author} {\bibinfo {author} {\bibfnamefont {R.}~\bibnamefont
			{Nandkishore}}\ and\ \bibinfo {author} {\bibfnamefont {D.~A.}\ \bibnamefont
			{Huse}},\ }\bibfield  {title} {\bibinfo {title} {{Many-Body Localization and
				Thermalization in Quantum Statistical Mechanics}},\ }\href
	{https://doi.org/10.1146/annurev-conmatphys-031214-014726} {\bibfield
		{journal} {\bibinfo  {journal} {Annual Review of Condensed Matter Physics}\
		}\textbf {\bibinfo {volume} {6}},\ \bibinfo {pages} {15} (\bibinfo {year}
		{2015})}\BibitemShut {NoStop}%
	\bibitem [{\citenamefont {{Landau}}\ and\ \citenamefont
		{{Lifshitz}}(1969)}]{LandauStatPhys}%
	\BibitemOpen
	\bibfield  {author} {\bibinfo {author} {\bibfnamefont {L.~D.}\ \bibnamefont
			{{Landau}}}\ and\ \bibinfo {author} {\bibfnamefont {E.~M.}\ \bibnamefont
			{{Lifshitz}}},\ }\href@noop {} {\emph {\bibinfo {title} {{Statistical
					physics. Pt.1}}}}\ (\bibinfo {year} {1969})\BibitemShut {NoStop}%
	\bibitem [{\citenamefont {Breuer}\ and\ \citenamefont
		{Petruccione}(2002)}]{breuer2002theory}%
	\BibitemOpen
	\bibfield  {author} {\bibinfo {author} {\bibfnamefont {H.-P.}\ \bibnamefont
			{Breuer}}\ and\ \bibinfo {author} {\bibfnamefont {F.}~\bibnamefont
			{Petruccione}},\ }\href@noop {} {\emph {\bibinfo {title} {The theory of open
				quantum systems}}}\ (\bibinfo  {publisher} {Oxford University Press},\
	\bibinfo {year} {2002})\BibitemShut {NoStop}%
	\bibitem [{\citenamefont {Micklitz}\ \emph {et~al.}(2022)\citenamefont
		{Micklitz}, \citenamefont {Morningstar}, \citenamefont {Altland},\ and\
		\citenamefont {Huse}}]{micklitz2022emergence}%
	\BibitemOpen
	\bibfield  {author} {\bibinfo {author} {\bibfnamefont {T.}~\bibnamefont
			{Micklitz}}, \bibinfo {author} {\bibfnamefont {A.}~\bibnamefont
			{Morningstar}}, \bibinfo {author} {\bibfnamefont {A.}~\bibnamefont
			{Altland}},\ and\ \bibinfo {author} {\bibfnamefont {D.~A.}\ \bibnamefont
			{Huse}},\ }\bibfield  {title} {\bibinfo {title} {Emergence of fermi's golden
			rule},\ }\href {https://doi.org/10.1103/PhysRevLett.129.140402} {\bibfield
		{journal} {\bibinfo  {journal} {Phys. Rev. Lett.}\ }\textbf {\bibinfo
			{volume} {129}},\ \bibinfo {pages} {140402} (\bibinfo {year}
		{2022})}\BibitemShut {NoStop}%
	\bibitem [{\citenamefont {Georges}\ \emph {et~al.}(1996)\citenamefont
		{Georges}, \citenamefont {Kotliar}, \citenamefont {Krauth},\ and\
		\citenamefont {Rozenberg}}]{Georges1996}%
	\BibitemOpen
	\bibfield  {author} {\bibinfo {author} {\bibfnamefont {A.}~\bibnamefont
			{Georges}}, \bibinfo {author} {\bibfnamefont {G.}~\bibnamefont {Kotliar}},
		\bibinfo {author} {\bibfnamefont {W.}~\bibnamefont {Krauth}},\ and\ \bibinfo
		{author} {\bibfnamefont {M.~J.}\ \bibnamefont {Rozenberg}},\ }\bibfield
	{title} {\bibinfo {title} {Dynamical mean-field theory of strongly correlated
			fermion systems and the limit of infinite dimensions},\ }\href
	{https://doi.org/10.1103/RevModPhys.68.13} {\bibfield  {journal} {\bibinfo
			{journal} {Rev. Mod. Phys.}\ }\textbf {\bibinfo {volume} {68}},\ \bibinfo
		{pages} {13} (\bibinfo {year} {1996})}\BibitemShut {NoStop}%
	\bibitem [{\citenamefont {Aoki}\ \emph {et~al.}(2014)\citenamefont {Aoki},
		\citenamefont {Tsuji}, \citenamefont {Eckstein}, \citenamefont {Kollar},
		\citenamefont {Oka},\ and\ \citenamefont {Werner}}]{Aoki2014}%
	\BibitemOpen
	\bibfield  {author} {\bibinfo {author} {\bibfnamefont {H.}~\bibnamefont
			{Aoki}}, \bibinfo {author} {\bibfnamefont {N.}~\bibnamefont {Tsuji}},
		\bibinfo {author} {\bibfnamefont {M.}~\bibnamefont {Eckstein}}, \bibinfo
		{author} {\bibfnamefont {M.}~\bibnamefont {Kollar}}, \bibinfo {author}
		{\bibfnamefont {T.}~\bibnamefont {Oka}},\ and\ \bibinfo {author}
		{\bibfnamefont {P.}~\bibnamefont {Werner}},\ }\bibfield  {title} {\bibinfo
		{title} {Nonequilibrium dynamical mean-field theory and its applications},\
	}\href {https://doi.org/10.1103/RevModPhys.86.779} {\bibfield  {journal}
		{\bibinfo  {journal} {Rev. Mod. Phys.}\ }\textbf {\bibinfo {volume} {86}},\
		\bibinfo {pages} {779} (\bibinfo {year} {2014})}\BibitemShut {NoStop}%
	\bibitem [{\citenamefont {Brunet}\ and\ \citenamefont
		{Derrida}(2015)}]{brunetderrida2015}%
	\BibitemOpen
	\bibfield  {author} {\bibinfo {author} {\bibfnamefont {{\'E}.}~\bibnamefont
			{Brunet}}\ and\ \bibinfo {author} {\bibfnamefont {B.}~\bibnamefont
			{Derrida}},\ }\bibfield  {title} {\bibinfo {title} {An exactly solvable
			travelling wave equation in the fisher--kpp class},\ }\href
	{https://doi.org/10.1007/s10955-015-1350-6} {\bibfield  {journal} {\bibinfo
			{journal} {Journal of Statistical Physics}\ }\textbf {\bibinfo {volume}
			{161}},\ \bibinfo {pages} {801} (\bibinfo {year} {2015})}\BibitemShut
	{NoStop}%
	\bibitem [{\citenamefont {Aleiner}\ \emph {et~al.}(2016)\citenamefont
		{Aleiner}, \citenamefont {Faoro},\ and\ \citenamefont
		{Ioffe}}]{ALEINER2016378}%
	\BibitemOpen
	\bibfield  {author} {\bibinfo {author} {\bibfnamefont {I.~L.}\ \bibnamefont
			{Aleiner}}, \bibinfo {author} {\bibfnamefont {L.}~\bibnamefont {Faoro}},\
		and\ \bibinfo {author} {\bibfnamefont {L.~B.}\ \bibnamefont {Ioffe}},\
	}\bibfield  {title} {\bibinfo {title} {Microscopic model of quantum butterfly
			effect: Out-of-time-order correlators and traveling combustion waves},\
	}\href {https://doi.org/https://doi.org/10.1016/j.aop.2016.09.006} {\bibfield
		{journal} {\bibinfo  {journal} {Annals of Physics}\ }\textbf {\bibinfo
			{volume} {375}},\ \bibinfo {pages} {378} (\bibinfo {year}
		{2016})}\BibitemShut {NoStop}%
	\bibitem [{\citenamefont {Xu}\ and\ \citenamefont
		{Swingle}(2019)}]{xu2019locality}%
	\BibitemOpen
	\bibfield  {author} {\bibinfo {author} {\bibfnamefont {S.}~\bibnamefont
			{Xu}}\ and\ \bibinfo {author} {\bibfnamefont {B.}~\bibnamefont {Swingle}},\
	}\bibfield  {title} {\bibinfo {title} {Locality, quantum fluctuations, and
			scrambling},\ }\href {https://doi.org/10.1103/PhysRevX.9.031048} {\bibfield
		{journal} {\bibinfo  {journal} {Phys. Rev. X}\ }\textbf {\bibinfo {volume}
			{9}},\ \bibinfo {pages} {031048} (\bibinfo {year} {2019})}\BibitemShut
	{NoStop}%
	\bibitem [{\citenamefont {Aron}\ \emph
		{et~al.}(2023{\natexlab{a}})\citenamefont {Aron}, \citenamefont {Éric
			Brunet},\ and\ \citenamefont {Mitra}}]{aron2023traveling}%
	\BibitemOpen
	\bibfield  {author} {\bibinfo {author} {\bibfnamefont {C.}~\bibnamefont
			{Aron}}, \bibinfo {author} {\bibnamefont {Éric Brunet}},\ and\ \bibinfo
		{author} {\bibfnamefont {A.}~\bibnamefont {Mitra}},\ }\bibfield  {title}
	{\bibinfo {title} {{Traveling discontinuity at the quantum butterfly
				front}},\ }\href {https://doi.org/10.21468/SciPostPhys.15.2.042} {\bibfield
		{journal} {\bibinfo  {journal} {SciPost Phys.}\ }\textbf {\bibinfo {volume}
			{15}},\ \bibinfo {pages} {042} (\bibinfo {year}
		{2023}{\natexlab{a}})}\BibitemShut {NoStop}%
	\bibitem [{\citenamefont {Aron}\ \emph
		{et~al.}(2023{\natexlab{b}})\citenamefont {Aron}, \citenamefont {Brunet},\
		and\ \citenamefont {Mitra}}]{aron2023kinetics}%
	\BibitemOpen
	\bibfield  {author} {\bibinfo {author} {\bibfnamefont {C.}~\bibnamefont
			{Aron}}, \bibinfo {author} {\bibfnamefont {E.}~\bibnamefont {Brunet}},\ and\
		\bibinfo {author} {\bibfnamefont {A.}~\bibnamefont {Mitra}},\ }\bibfield
	{title} {\bibinfo {title} {Kinetics of information scrambling in correlated
			metals: Disorder-driven transition from shock wave to fisher or
			kolmogorov-petrovsky-piskunov dynamics},\ }\href
	{https://doi.org/10.1103/PhysRevB.108.L241106} {\bibfield  {journal}
		{\bibinfo  {journal} {Phys. Rev. B}\ }\textbf {\bibinfo {volume} {108}},\
		\bibinfo {pages} {L241106} (\bibinfo {year}
		{2023}{\natexlab{b}})}\BibitemShut {NoStop}%
	\bibitem [{sup()}]{supplementary}%
	\BibitemOpen
	\href@noop {} {\bibinfo {title} {Supplemental material, where we detail the
			excitation protocol.}}\BibitemShut {Stop}%
	\bibitem [{\citenamefont {Eckstein}\ \emph {et~al.}(2009)\citenamefont
		{Eckstein}, \citenamefont {Kollar},\ and\ \citenamefont
		{Werner}}]{eckstein2009thermalization}%
	\BibitemOpen
	\bibfield  {author} {\bibinfo {author} {\bibfnamefont {M.}~\bibnamefont
			{Eckstein}}, \bibinfo {author} {\bibfnamefont {M.}~\bibnamefont {Kollar}},\
		and\ \bibinfo {author} {\bibfnamefont {P.}~\bibnamefont {Werner}},\
	}\bibfield  {title} {\bibinfo {title} {Thermalization after an interaction
			quench in the hubbard model},\ }\href
	{https://doi.org/10.1103/PhysRevLett.103.056403} {\bibfield  {journal}
		{\bibinfo  {journal} {Phys. Rev. Lett.}\ }\textbf {\bibinfo {volume} {103}},\
		\bibinfo {pages} {056403} (\bibinfo {year} {2009})}\BibitemShut {NoStop}%
	\bibitem [{\citenamefont {Peronaci}\ \emph {et~al.}(2018)\citenamefont
		{Peronaci}, \citenamefont {Schir\'o},\ and\ \citenamefont
		{Parcollet}}]{peronaci2018resonant}%
	\BibitemOpen
	\bibfield  {author} {\bibinfo {author} {\bibfnamefont {F.}~\bibnamefont
			{Peronaci}}, \bibinfo {author} {\bibfnamefont {M.}~\bibnamefont {Schir\'o}},\
		and\ \bibinfo {author} {\bibfnamefont {O.}~\bibnamefont {Parcollet}},\
	}\bibfield  {title} {\bibinfo {title} {Resonant thermalization of
			periodically driven strongly correlated electrons},\ }\href
	{https://doi.org/10.1103/PhysRevLett.120.197601} {\bibfield  {journal}
		{\bibinfo  {journal} {Phys. Rev. Lett.}\ }\textbf {\bibinfo {volume} {120}},\
		\bibinfo {pages} {197601} (\bibinfo {year} {2018})}\BibitemShut {NoStop}%
	\bibitem [{\citenamefont {Florens}(2007)}]{florens2007nanoscale}%
	\BibitemOpen
	\bibfield  {author} {\bibinfo {author} {\bibfnamefont {S.}~\bibnamefont
			{Florens}},\ }\bibfield  {title} {\bibinfo {title} {Nanoscale dynamical
			mean-field theory for molecules and mesoscopic devices in the
			strong-correlation regime},\ }\href
	{https://doi.org/10.1103/PhysRevLett.99.046402} {\bibfield  {journal}
		{\bibinfo  {journal} {Phys. Rev. Lett.}\ }\textbf {\bibinfo {volume} {99}},\
		\bibinfo {pages} {046402} (\bibinfo {year} {2007})}\BibitemShut {NoStop}%
	\bibitem [{\citenamefont {Held}\ \emph {et~al.}(2013)\citenamefont {Held},
		\citenamefont {Peters},\ and\ \citenamefont {Toschi}}]{held2013poor}%
	\BibitemOpen
	\bibfield  {author} {\bibinfo {author} {\bibfnamefont {K.}~\bibnamefont
			{Held}}, \bibinfo {author} {\bibfnamefont {R.}~\bibnamefont {Peters}},\ and\
		\bibinfo {author} {\bibfnamefont {A.}~\bibnamefont {Toschi}},\ }\bibfield
	{title} {\bibinfo {title} {Poor man's understanding of kinks originating from
			strong electronic correlations},\ }\href
	{https://doi.org/10.1103/PhysRevLett.110.246402} {\bibfield  {journal}
		{\bibinfo  {journal} {Phys. Rev. Lett.}\ }\textbf {\bibinfo {volume} {110}},\
		\bibinfo {pages} {246402} (\bibinfo {year} {2013})}\BibitemShut {NoStop}%
	\bibitem [{\citenamefont {Georges}\ and\ \citenamefont
		{Krauth}(1992)}]{georges1992numerical}%
	\BibitemOpen
	\bibfield  {author} {\bibinfo {author} {\bibfnamefont {A.}~\bibnamefont
			{Georges}}\ and\ \bibinfo {author} {\bibfnamefont {W.}~\bibnamefont
			{Krauth}},\ }\bibfield  {title} {\bibinfo {title} {Numerical solution of the
			d=\ensuremath{\infty} hubbard model: Evidence for a mott transition},\ }\href
	{https://doi.org/10.1103/PhysRevLett.69.1240} {\bibfield  {journal} {\bibinfo
			{journal} {Phys. Rev. Lett.}\ }\textbf {\bibinfo {volume} {69}},\ \bibinfo
		{pages} {1240} (\bibinfo {year} {1992})}\BibitemShut {NoStop}%
	\bibitem [{\citenamefont {Rozenberg}\ \emph {et~al.}(1992)\citenamefont
		{Rozenberg}, \citenamefont {Zhang},\ and\ \citenamefont
		{Kotliar}}]{rozenberg1992mott}%
	\BibitemOpen
	\bibfield  {author} {\bibinfo {author} {\bibfnamefont {M.~J.}\ \bibnamefont
			{Rozenberg}}, \bibinfo {author} {\bibfnamefont {X.~Y.}\ \bibnamefont
			{Zhang}},\ and\ \bibinfo {author} {\bibfnamefont {G.}~\bibnamefont
			{Kotliar}},\ }\bibfield  {title} {\bibinfo {title} {Mott-hubbard transition
			in infinite dimensions},\ }\href
	{https://doi.org/10.1103/PhysRevLett.69.1236} {\bibfield  {journal} {\bibinfo
			{journal} {Phys. Rev. Lett.}\ }\textbf {\bibinfo {volume} {69}},\ \bibinfo
		{pages} {1236} (\bibinfo {year} {1992})}\BibitemShut {NoStop}%
	\bibitem [{\citenamefont {Rosch}\ \emph {et~al.}(2001)\citenamefont {Rosch},
		\citenamefont {Kroha},\ and\ \citenamefont {W\"olfle}}]{rosch2001kondo}%
	\BibitemOpen
	\bibfield  {author} {\bibinfo {author} {\bibfnamefont {A.}~\bibnamefont
			{Rosch}}, \bibinfo {author} {\bibfnamefont {J.}~\bibnamefont {Kroha}},\ and\
		\bibinfo {author} {\bibfnamefont {P.}~\bibnamefont {W\"olfle}},\ }\bibfield
	{title} {\bibinfo {title} {Kondo effect in quantum dots at high voltage:
			Universality and scaling},\ }\href
	{https://doi.org/10.1103/PhysRevLett.87.156802} {\bibfield  {journal}
		{\bibinfo  {journal} {Phys. Rev. Lett.}\ }\textbf {\bibinfo {volume} {87}},\
		\bibinfo {pages} {156802} (\bibinfo {year} {2001})}\BibitemShut {NoStop}%
	\bibitem [{\citenamefont {Kehrein}(2005)}]{kehrein2005scaling}%
	\BibitemOpen
	\bibfield  {author} {\bibinfo {author} {\bibfnamefont {S.}~\bibnamefont
			{Kehrein}},\ }\bibfield  {title} {\bibinfo {title} {Scaling and decoherence
			in the nonequilibrium kondo model},\ }\href
	{https://doi.org/10.1103/PhysRevLett.95.056602} {\bibfield  {journal}
		{\bibinfo  {journal} {Phys. Rev. Lett.}\ }\textbf {\bibinfo {volume} {95}},\
		\bibinfo {pages} {056602} (\bibinfo {year} {2005})}\BibitemShut {NoStop}%
	\bibitem [{\citenamefont {Mitra}\ \emph {et~al.}(2006)\citenamefont {Mitra},
		\citenamefont {Takei}, \citenamefont {Kim},\ and\ \citenamefont
		{Millis}}]{mitra2006nonequilibrium}%
	\BibitemOpen
	\bibfield  {author} {\bibinfo {author} {\bibfnamefont {A.}~\bibnamefont
			{Mitra}}, \bibinfo {author} {\bibfnamefont {S.}~\bibnamefont {Takei}},
		\bibinfo {author} {\bibfnamefont {Y.~B.}\ \bibnamefont {Kim}},\ and\ \bibinfo
		{author} {\bibfnamefont {A.~J.}\ \bibnamefont {Millis}},\ }\bibfield  {title}
	{\bibinfo {title} {Nonequilibrium quantum criticality in open electronic
			systems},\ }\href {https://doi.org/10.1103/PhysRevLett.97.236808} {\bibfield
		{journal} {\bibinfo  {journal} {Phys. Rev. Lett.}\ }\textbf {\bibinfo
			{volume} {97}},\ \bibinfo {pages} {236808} (\bibinfo {year}
		{2006})}\BibitemShut {NoStop}%
	\bibitem [{\citenamefont {Schir\'o}\ and\ \citenamefont
		{Mitra}(2014)}]{schiro2014transient}%
	\BibitemOpen
	\bibfield  {author} {\bibinfo {author} {\bibfnamefont {M.}~\bibnamefont
			{Schir\'o}}\ and\ \bibinfo {author} {\bibfnamefont {A.}~\bibnamefont
			{Mitra}},\ }\bibfield  {title} {\bibinfo {title} {Transient orthogonality
			catastrophe in a time-dependent nonequilibrium environment},\ }\href
	{https://doi.org/10.1103/PhysRevLett.112.246401} {\bibfield  {journal}
		{\bibinfo  {journal} {Phys. Rev. Lett.}\ }\textbf {\bibinfo {volume} {112}},\
		\bibinfo {pages} {246401} (\bibinfo {year} {2014})}\BibitemShut {NoStop}%
	\bibitem [{\citenamefont {Picano}\ \emph {et~al.}(2021)\citenamefont {Picano},
		\citenamefont {Li},\ and\ \citenamefont {Eckstein}}]{Picano2021}%
	\BibitemOpen
	\bibfield  {author} {\bibinfo {author} {\bibfnamefont {A.}~\bibnamefont
			{Picano}}, \bibinfo {author} {\bibfnamefont {J.}~\bibnamefont {Li}},\ and\
		\bibinfo {author} {\bibfnamefont {M.}~\bibnamefont {Eckstein}},\ }\bibfield
	{title} {\bibinfo {title} {Quantum boltzmann equation for strongly correlated
			electrons},\ }\href {https://doi.org/10.1103/PhysRevB.104.085108} {\bibfield
		{journal} {\bibinfo  {journal} {Phys. Rev. B}\ }\textbf {\bibinfo {volume}
			{104}},\ \bibinfo {pages} {085108} (\bibinfo {year} {2021})}\BibitemShut
	{NoStop}%
	\bibitem [{\citenamefont {Strand}\ \emph {et~al.}(2015)\citenamefont {Strand},
		\citenamefont {Eckstein},\ and\ \citenamefont
		{Werner}}]{strand2015nonequilibrium}%
	\BibitemOpen
	\bibfield  {author} {\bibinfo {author} {\bibfnamefont {H.~U.~R.}\
			\bibnamefont {Strand}}, \bibinfo {author} {\bibfnamefont {M.}~\bibnamefont
			{Eckstein}},\ and\ \bibinfo {author} {\bibfnamefont {P.}~\bibnamefont
			{Werner}},\ }\bibfield  {title} {\bibinfo {title} {Nonequilibrium dynamical
			mean-field theory for bosonic lattice models},\ }\href
	{https://doi.org/10.1103/PhysRevX.5.011038} {\bibfield  {journal} {\bibinfo
			{journal} {Phys. Rev. X}\ }\textbf {\bibinfo {volume} {5}},\ \bibinfo {pages}
		{011038} (\bibinfo {year} {2015})}\BibitemShut {NoStop}%
	\bibitem [{\citenamefont {Scarlatella}\ \emph {et~al.}(2021)\citenamefont
		{Scarlatella}, \citenamefont {Clerk}, \citenamefont {Fazio},\ and\
		\citenamefont {Schir\'o}}]{scarlatella2021dynamical}%
	\BibitemOpen
	\bibfield  {author} {\bibinfo {author} {\bibfnamefont {O.}~\bibnamefont
			{Scarlatella}}, \bibinfo {author} {\bibfnamefont {A.~A.}\ \bibnamefont
			{Clerk}}, \bibinfo {author} {\bibfnamefont {R.}~\bibnamefont {Fazio}},\ and\
		\bibinfo {author} {\bibfnamefont {M.}~\bibnamefont {Schir\'o}},\ }\bibfield
	{title} {\bibinfo {title} {Dynamical mean-field theory for markovian open
			quantum many-body systems},\ }\href
	{https://doi.org/10.1103/PhysRevX.11.031018} {\bibfield  {journal} {\bibinfo
			{journal} {Phys. Rev. X}\ }\textbf {\bibinfo {volume} {11}},\ \bibinfo
		{pages} {031018} (\bibinfo {year} {2021})}\BibitemShut {NoStop}%
	\bibitem [{\citenamefont {Shih}\ and\ \citenamefont
		{Berkelbach}(2022)}]{shih2022anharmonic}%
	\BibitemOpen
	\bibfield  {author} {\bibinfo {author} {\bibfnamefont {P.}~\bibnamefont
			{Shih}}\ and\ \bibinfo {author} {\bibfnamefont {T.~C.}\ \bibnamefont
			{Berkelbach}},\ }\bibfield  {title} {\bibinfo {title} {Anharmonic lattice
			dynamics from vibrational dynamical mean-field theory},\ }\href
	{https://doi.org/10.1103/PhysRevB.106.144307} {\bibfield  {journal} {\bibinfo
			{journal} {Phys. Rev. B}\ }\textbf {\bibinfo {volume} {106}},\ \bibinfo
		{pages} {144307} (\bibinfo {year} {2022})}\BibitemShut {NoStop}%
	\bibitem [{\citenamefont {Miranda}\ and\ \citenamefont
		{Dobrosavljević}(2005)}]{Miranda_2005}%
	\BibitemOpen
	\bibfield  {author} {\bibinfo {author} {\bibfnamefont {E.}~\bibnamefont
			{Miranda}}\ and\ \bibinfo {author} {\bibfnamefont {V.}~\bibnamefont
			{Dobrosavljević}},\ }\bibfield  {title} {\bibinfo {title} {Disorder-driven
			non-fermi liquid behaviour of correlated electrons},\ }\href
	{https://doi.org/10.1088/0034-4885/68/10/R02} {\bibfield  {journal} {\bibinfo
			{journal} {Reports on Progress in Physics}\ }\textbf {\bibinfo {volume}
			{68}},\ \bibinfo {pages} {2337} (\bibinfo {year} {2005})}\BibitemShut
	{NoStop}%
\end{thebibliography}

\begin{thebibliography}{8}%
	\makeatletter
	\providecommand \@ifxundefined [1]{%
		\@ifx{#1\undefined}
	}%
	\providecommand \@ifnum [1]{%
		\ifnum #1\expandafter \@firstoftwo
		\else \expandafter \@secondoftwo
		\fi
	}%
	\providecommand \@ifx [1]{%
		\ifx #1\expandafter \@firstoftwo
		\else \expandafter \@secondoftwo
		\fi
	}%
	\providecommand \natexlab [1]{#1}%
	\providecommand \enquote  [1]{``#1''}%
	\providecommand \bibnamefont  [1]{#1}%
	\providecommand \bibfnamefont [1]{#1}%
	\providecommand \citenamefont [1]{#1}%
	\providecommand \href@noop [0]{\@secondoftwo}%
	\providecommand \href [0]{\begingroup \@sanitize@url \@href}%
	\providecommand \@href[1]{\@@startlink{#1}\@@href}%
	\providecommand \@@href[1]{\endgroup#1\@@endlink}%
	\providecommand \@sanitize@url [0]{\catcode `\\12\catcode `\$12\catcode
		`\&12\catcode `\#12\catcode `\^12\catcode `\_12\catcode `\%12\relax}%
	\providecommand \@@startlink[1]{}%
	\providecommand \@@endlink[0]{}%
	\providecommand \url  [0]{\begingroup\@sanitize@url \@url }%
	\providecommand \@url [1]{\endgroup\@href {#1}{\urlprefix }}%
	\providecommand \urlprefix  [0]{URL }%
	\providecommand \Eprint [0]{\href }%
	\providecommand \doibase [0]{https://doi.org/}%
	\providecommand \selectlanguage [0]{\@gobble}%
	\providecommand \bibinfo  [0]{\@secondoftwo}%
	\providecommand \bibfield  [0]{\@secondoftwo}%
	\providecommand \translation [1]{[#1]}%
	\providecommand \BibitemOpen [0]{}%
	\providecommand \bibitemStop [0]{}%
	\providecommand \bibitemNoStop [0]{.\EOS\space}%
	\providecommand \EOS [0]{\spacefactor3000\relax}%
	\providecommand \BibitemShut  [1]{\csname bibitem#1\endcsname}%
	\let\auto@bib@innerbib\@empty
	\bibitem [{\citenamefont {Georges}\ \emph {et~al.}(1996)\citenamefont
		{Georges}, \citenamefont {Kotliar}, \citenamefont {Krauth},\ and\
		\citenamefont {Rozenberg}}]{Georges1996}%
	\BibitemOpen
	\bibfield  {author} {\bibinfo {author} {\bibfnamefont {A.}~\bibnamefont
			{Georges}}, \bibinfo {author} {\bibfnamefont {G.}~\bibnamefont {Kotliar}},
		\bibinfo {author} {\bibfnamefont {W.}~\bibnamefont {Krauth}},\ and\ \bibinfo
		{author} {\bibfnamefont {M.~J.}\ \bibnamefont {Rozenberg}},\ }\bibfield
	{title} {\bibinfo {title} {Dynamical mean-field theory of strongly correlated
			fermion systems and the limit of infinite dimensions},\ }\href
	{https://doi.org/10.1103/RevModPhys.68.13} {\bibfield  {journal} {\bibinfo
			{journal} {Rev. Mod. Phys.}\ }\textbf {\bibinfo {volume} {68}},\ \bibinfo
		{pages} {13} (\bibinfo {year} {1996})}\BibitemShut {NoStop}%
	\bibitem [{\citenamefont {Aoki}\ \emph {et~al.}(2014)\citenamefont {Aoki},
		\citenamefont {Tsuji}, \citenamefont {Eckstein}, \citenamefont {Kollar},
		\citenamefont {Oka},\ and\ \citenamefont {Werner}}]{Aoki2014}%
	\BibitemOpen
	\bibfield  {author} {\bibinfo {author} {\bibfnamefont {H.}~\bibnamefont
			{Aoki}}, \bibinfo {author} {\bibfnamefont {N.}~\bibnamefont {Tsuji}},
		\bibinfo {author} {\bibfnamefont {M.}~\bibnamefont {Eckstein}}, \bibinfo
		{author} {\bibfnamefont {M.}~\bibnamefont {Kollar}}, \bibinfo {author}
		{\bibfnamefont {T.}~\bibnamefont {Oka}},\ and\ \bibinfo {author}
		{\bibfnamefont {P.}~\bibnamefont {Werner}},\ }\bibfield  {title} {\bibinfo
		{title} {Nonequilibrium dynamical mean-field theory and its applications},\
	}\href {https://doi.org/10.1103/RevModPhys.86.779} {\bibfield  {journal}
		{\bibinfo  {journal} {Rev. Mod. Phys.}\ }\textbf {\bibinfo {volume} {86}},\
		\bibinfo {pages} {779} (\bibinfo {year} {2014})}\BibitemShut {NoStop}%
	\bibitem [{\citenamefont {Schüler}\ \emph {et~al.}(2020)\citenamefont
		{Schüler}, \citenamefont {Gole{\v{z}}}, \citenamefont {Murakami},
		\citenamefont {Bittner}, \citenamefont {Herrmann}, \citenamefont {Strand},
		\citenamefont {Werner},\ and\ \citenamefont {Eckstein}}]{Schuler2020}%
	\BibitemOpen
	\bibfield  {author} {\bibinfo {author} {\bibfnamefont {M.}~\bibnamefont
			{Schüler}}, \bibinfo {author} {\bibfnamefont {D.}~\bibnamefont
			{Gole{\v{z}}}}, \bibinfo {author} {\bibfnamefont {Y.}~\bibnamefont
			{Murakami}}, \bibinfo {author} {\bibfnamefont {N.}~\bibnamefont {Bittner}},
		\bibinfo {author} {\bibfnamefont {A.}~\bibnamefont {Herrmann}}, \bibinfo
		{author} {\bibfnamefont {H.~U.}\ \bibnamefont {Strand}}, \bibinfo {author}
		{\bibfnamefont {P.}~\bibnamefont {Werner}},\ and\ \bibinfo {author}
		{\bibfnamefont {M.}~\bibnamefont {Eckstein}},\ }\bibfield  {title} {\bibinfo
		{title} {{NESSi}: The non-equilibrium systems simulation package},\ }\href
	{https://doi.org/10.1016/j.cpc.2020.107484} {\bibfield  {journal} {\bibinfo
			{journal} {Computer Physics Communications}\ }\textbf {\bibinfo {volume}
			{257}},\ \bibinfo {pages} {107484} (\bibinfo {year} {2020})}\BibitemShut
	{NoStop}%
	\bibitem [{\citenamefont {Eckstein}\ \emph {et~al.}(2010)\citenamefont
		{Eckstein}, \citenamefont {Kollar},\ and\ \citenamefont
		{Werner}}]{Eckstein2010}%
	\BibitemOpen
	\bibfield  {author} {\bibinfo {author} {\bibfnamefont {M.}~\bibnamefont
			{Eckstein}}, \bibinfo {author} {\bibfnamefont {M.}~\bibnamefont {Kollar}},\
		and\ \bibinfo {author} {\bibfnamefont {P.}~\bibnamefont {Werner}},\
	}\bibfield  {title} {\bibinfo {title} {{Interaction quench in the Hubbard
				model: Relaxation of the spectral function and the optical conductivity}},\
	}\href {https://doi.org/10.1103/PhysRevB.81.115131} {\bibfield  {journal}
		{\bibinfo  {journal} {Physical Review B}\ }\textbf {\bibinfo {volume} {8}},\
		\bibinfo {pages} {1} (\bibinfo {year} {2010})}\BibitemShut {NoStop}%
	\bibitem [{\citenamefont {Picano}\ \emph {et~al.}(2023)\citenamefont {Picano},
		\citenamefont {Grandi},\ and\ \citenamefont
		{Eckstein}}]{Picano2023_inhomogeneous}%
	\BibitemOpen
	\bibfield  {author} {\bibinfo {author} {\bibfnamefont {A.}~\bibnamefont
			{Picano}}, \bibinfo {author} {\bibfnamefont {F.}~\bibnamefont {Grandi}},\
		and\ \bibinfo {author} {\bibfnamefont {M.}~\bibnamefont {Eckstein}},\
	}\bibfield  {title} {\bibinfo {title} {Inhomogeneous disordering at a
			photoinduced charge density wave transition},\ }\href
	{https://doi.org/10.1103/PhysRevB.107.245112} {\bibfield  {journal} {\bibinfo
			{journal} {Phys. Rev. B}\ }\textbf {\bibinfo {volume} {107}},\ \bibinfo
		{pages} {245112} (\bibinfo {year} {2023})}\BibitemShut {NoStop}%
	\bibitem [{\citenamefont {Picano}\ \emph {et~al.}(2021)\citenamefont {Picano},
		\citenamefont {Li},\ and\ \citenamefont {Eckstein}}]{Picano2021}%
	\BibitemOpen
	\bibfield  {author} {\bibinfo {author} {\bibfnamefont {A.}~\bibnamefont
			{Picano}}, \bibinfo {author} {\bibfnamefont {J.}~\bibnamefont {Li}},\ and\
		\bibinfo {author} {\bibfnamefont {M.}~\bibnamefont {Eckstein}},\ }\bibfield
	{title} {\bibinfo {title} {Quantum boltzmann equation for strongly correlated
			electrons},\ }\href {https://doi.org/10.1103/PhysRevB.104.085108} {\bibfield
		{journal} {\bibinfo  {journal} {Phys. Rev. B}\ }\textbf {\bibinfo {volume}
			{104}},\ \bibinfo {pages} {085108} (\bibinfo {year} {2021})}\BibitemShut
	{NoStop}%
	\bibitem [{\citenamefont {Brunet}\ and\ \citenamefont
		{Derrida}(2015)}]{brunetderrida2015}%
	\BibitemOpen
	\bibfield  {author} {\bibinfo {author} {\bibfnamefont {{\'E}.}~\bibnamefont
			{Brunet}}\ and\ \bibinfo {author} {\bibfnamefont {B.}~\bibnamefont
			{Derrida}},\ }\bibfield  {title} {\bibinfo {title} {An exactly solvable
			travelling wave equation in the fisher--kpp class},\ }\href
	{https://doi.org/10.1007/s10955-015-1350-6} {\bibfield  {journal} {\bibinfo
			{journal} {Journal of Statistical Physics}\ }\textbf {\bibinfo {volume}
			{161}},\ \bibinfo {pages} {801} (\bibinfo {year} {2015})}\BibitemShut
	{NoStop}%
	\bibitem [{\citenamefont {Aron}\ \emph {et~al.}(2023)\citenamefont {Aron},
		\citenamefont {Brunet},\ and\ \citenamefont {Mitra}}]{aron2023kinetics}%
	\BibitemOpen
	\bibfield  {author} {\bibinfo {author} {\bibfnamefont {C.}~\bibnamefont
			{Aron}}, \bibinfo {author} {\bibfnamefont {E.}~\bibnamefont {Brunet}},\ and\
		\bibinfo {author} {\bibfnamefont {A.}~\bibnamefont {Mitra}},\ }\bibfield
	{title} {\bibinfo {title} {Kinetics of information scrambling in correlated
			metals: Disorder-driven transition from shock wave to fisher or
			kolmogorov-petrovsky-piskunov dynamics},\ }\href
	{https://doi.org/10.1103/PhysRevB.108.L241106} {\bibfield  {journal}
		{\bibinfo  {journal} {Phys. Rev. B}\ }\textbf {\bibinfo {volume} {108}},\
		\bibinfo {pages} {L241106} (\bibinfo {year} {2023})}\BibitemShut {NoStop}%
\end{thebibliography}
%

\end{document}